\documentclass[apj,twocolumn,twocolappendix,numberedappendix]{openjournal}
\usepackage{amsmath}
\usepackage{booktabs}
\usepackage{multirow}
\usepackage{color}
\usepackage{soul}
\usepackage{threeparttable}
\usepackage{float}
\usepackage{graphicx}
\usepackage{CJK}
\usepackage{xspace}
\usepackage{afterpage}
\usepackage{placeins}
\usepackage[breaklinks,colorlinks,citecolor=blue,urlcolor=blue,linkcolor=blue,filecolor=blue]{hyperref}
\usepackage{xcolor}
\newcommand{\refrep}[1]{#1}

\shorttitle{The Mirage or Miracle Survey: A Remarkably Luminous Galaxy at $z_{\rm{spec}}=14.44$}
\shortauthors{Naidu et al.}

\usepackage[export]{adjustbox}

\newcommand{\orcidauthor}[3]{\author{\href{http://orcid.org/#1}{#2$^{#3}$}}}
\newcommand{\nion}[2]{#1\,\textsc{#2}}

\begin{document}
\begin{CJK*}{UTF8}{gbsn}

\title{\vspace{-1cm} A Cosmic Miracle: A Remarkably Luminous Galaxy at $\MakeLowercase{z}_{\MakeLowercase{\rm{spec}}}=14.44$ Confirmed with JWST \vspace{-1.8cm}}

\orcidauthor{0000-0003-3997-5705}{Rohan P. Naidu}{1, *,\dagger,\ddagger}
\orcidauthor{0000-0001-5851-6649}{Pascal A.\ Oesch}{2,3,4,\ddagger}
\orcidauthor{0000-0003-2680-005X}{Gabriel Brammer}{3,4}
\orcidauthor{0000-0001-8928-4465}{Andrea Weibel}{2}
\orcidauthor{0000-0002-0682-3310}{Yijia Li (李轶佳)}{5,6}
\orcidauthor{0000-0003-2871-127X}{Jorryt Matthee}{7}
\orcidauthor{0000-0002-0302-2577}{John Chisholm}{8}
\orcidauthor{0009-0001-2808-4918}{Clara L. Pollock}{3,4}
\orcidauthor{0000-0002-9389-7413}{Kasper E. Heintz}{3,4,2}
\orcidauthor{0000-0002-9280-7594}{Benjamin D. Johnson}{10}
\orcidauthor{0000-0002-6196-823X}{Xuejian Shen}{1}
\orcidauthor{0000-0002-4684-9005}{Raphael E. Hviding}{11}
\orcidauthor{0000-0001-6755-1315}{Joel Leja}{5,12,6}
\orcidauthor{0000-0002-8224-4505}{Sandro Tacchella}{13,14}
\orcidauthor{0009-0008-0444-4289}{Arpita Ganguly}{2}
\orcidauthor{0000-0002-1369-6452}{Callum Witten}{2}
\orcidauthor{0000-0002-7570-0824}{Hakim Atek}{15}
\orcidauthor{0000-0002-5615-6018}{Sirio Belli}{16}
\orcidauthor{0000-0002-0974-5266}{Sownak Bose}{17}
\orcidauthor{0000-0002-4989-2471}{Rychard Bouwens}{18}
\orcidauthor{0000-0001-8460-1564}{Pratika Dayal}{19}
\orcidauthor{0000-0002-2662-8803}{Roberto Decarli}{20}
\orcidauthor{0000-0002-2380-9801}{Anna de Graaff}{11}
\orcidauthor{0000-0001-7440-8832}{Yoshinobu Fudamoto}{21}
\orcidauthor{0009-0004-3835-0089}{Emma Giovinazzo}{2}
\orcidauthor{0000-0002-5612-3427}{Jenny E. Greene}{22}
\orcidauthor{0000-0002-8096-2837}{Garth Illingworth}{23}
\orcidauthor{0000-0002-7779-8677}{Akio K. Inoue}{24,25}
\orcidauthor{0000-0001-8411-1012}{Sarah G. Kane}{26}
\orcidauthor{0000-0002-2057-5376}{Ivo Labbe}{27}
\orcidauthor{0000-0002-5757-4334}{Ecaterina Leonova}{28, 29}
\orcidauthor{0000-0001-8442-1846}{Rui Marques-Chaves}{2}
\orcidauthor{0000-0001-5492-4522}{Romain A. Meyer}{2}
\orcidauthor{0000-0002-7524-374X}{Erica J. Nelson}{30}
\orcidauthor{0000-0002-4140-1367}{Guido Roberts-Borsani}{31}
\orcidauthor{0000-0001-7144-7182}{Daniel Schaerer}{2,32}
\orcidauthor{0000-0003-3769-9559}{Robert A. Simcoe}{1}
\orcidauthor{0000-0001-7768-5309}{Mauro Stefanon}{33,34}
\orcidauthor{0000-0001-6958-7856}{Yuma Sugahara}{24,25}
\orcidauthor{0000-0003-3631-7176}{Sune Toft}{3,4}
\orcidauthor{0000-0002-5027-0135}{Arjen van der Wel}{35}
\orcidauthor{0000-0002-8282-9888}{Pieter van Dokkum}{36}
\orcidauthor{0000-0003-4793-7880}{Fabian Walter}{11}
\orcidauthor{0000-0002-4465-8264}{Darach Watson}{3,4}
\orcidauthor{0000-0003-1614-196X}{John R. Weaver}{37}
\orcidauthor{0000-0001-7160-3632}{Katherine E. Whitaker}{37,3}

\affiliation{$^1$ MIT Kavli Institute for Astrophysics and Space Research, 70 Vassar Street, Cambridge, MA 02139, USA}
\affiliation{$^2$ Department of Astronomy, University of Geneva, Chemin Pegasi 51, 1290 Versoix, Switzerland}
\affiliation{$^3$ Cosmic Dawn Center (DAWN), Copenhagen, Denmark}
\affiliation{$^4$ Niels Bohr Institute, University of Copenhagen, Jagtvej 128, K{\o}benhavn N, DK-2200, Denmark}
\affiliation{$^5$ Department of Astronomy \& Astrophysics, The Pennsylvania State University, University Park, PA 16802, USA}
\affiliation{$^6$ Institute for Gravitation and the Cosmos, The Pennsylvania State University, University Park, PA 16802, USA}
\affiliation{$^7$ Institute of Science and Technology Austria (ISTA), Am Campus 1, 3400 Klosterneuburg, Austria}
\affiliation{$^8$ Department of Astronomy, The University of Texas at Austin, Austin, TX, USA}
\affiliation{$^{10}$ Center for Astrophysics $|$ Harvard \& Smithsonian, 60 Garden St., Cambridge MA 02138 USA}
\affiliation{$^{11}$ Max-Planck-Institut f\"ur Astronomie, K\"onigstuhl 17, D-69117 Heidelberg, Germany}
\affiliation{$^{12}$ Institute for Computational \& Data Sciences, The Pennsylvania State University, University Park, PA 16802, USA}
\affiliation{$^{13}$ The Kavli Institute for Cosmology (KICC), University of Cambridge, Madingley Road, Cambridge, CB3 0HA, UK}
\affiliation{$^{14}$ Cavendish Laboratory, University of Cambridge, 19 JJ Thomson Avenue, Cambridge, CB3 0HE, UK}
\affiliation{$^{15}$ Institut d’Astrophysique de Paris, CNRS, Sorbonne Universit\'e, 98bis Boulevard Arago, 75014, Paris, France}
\affiliation{$^{16}$ Dipartimento di Fisica e Astronomia, Universit\`a di Bologna, Bologna, Italy}
\affiliation{$^{17}$ Institute for Computational Cosmology, Department of Physics, Durham University, South Road, Durham DH1 3LE, UK}
\affiliation{$^{18}$ Leiden Observatory, Leiden University, P.O. Box 9513, NL-2300 RA Leiden, the Netherlands}
\affiliation{$^{19}$ Kapteyn Astronomical Institute, University of Groningen, P.O. Box 800, 9700 AV Groningen, The Netherlands}
\affiliation{$^{20}$ INAF \textendash{} Osservatorio di Astrofisica e Scienza dello Spazio di Bologna, via Gobetti 93/3, I-40129 Bologna, Italy}
\affiliation{$^{21}$ Center for Frontier Science, Chiba University, 1-33 Yayoi-cho, Inage-ku, Chiba 263-8522, Japan}
\affiliation{$^{22}$ Department of Astrophysical Sciences, Princeton University, Princeton, NJ 08544, USA}
\affiliation{$^{23}$ Department of Astronomy and Astrophysics, University of California, Santa Cruz, CA 95064, USA}
\affiliation{$^{24}$ Department of Physics, School of Advanced Science and Engineering, Faculty of Science and Engineering, Waseda University, 3-4-1 Okubo, Shinjuku, Tokyo 169-8555, Japan}
\affiliation{$^{25}$ Waseda Research Institute for Science and Engineering, Faculty of Science and Engineering, Waseda University, 3-4-1 Okubo, Shinjuku, Tokyo 169-8555, Japan}
\affiliation{$^{26}$ Institute of Astronomy, University of Cambridge, Madingley Road, Cambridge CB3 0HA, UK}
\affiliation{$^{27}$ Centre for Astrophysics and Supercomputing, Swinburne University of Technology, Melbourne, VIC 3122, Australia}
\affiliation{$^{28}$ GRAPPA, Anton Pannekoek Institute for Astronomy and Institute of High-Energy Physics}
\affiliation{$^{29}$ University of Amsterdam, Science Park 904, NL-1098 XH Amsterdam, the Netherlands}
\affiliation{$^{30}$ Department for Astrophysical and Planetary Science, University of Colorado, Boulder, CO 80309, USA}
\affiliation{$^{31}$ Department of Physics \& Astronomy, University College London, London, WC1E 6BT, UK}
\affiliation{$^{32}$ CNRS, IRAP, 14 Avenue E. Belin, 31400 Toulouse, France}
\affiliation{$^{33}$ Departament d'Astronomia i Astrof\`isica, Universitat de Val\`encia, C. Dr. Moliner 50, E-46100 Burjassot, Val\`encia, Spain}
\affiliation{$^{34}$ Unidad Asociada CSIC ``Grupo de Astrof\'isica Extragal\'actica y Cosmolog\'ia"}
\affiliation{$^{35}$ Sterrenkundig Observatorium, Universiteit Gent, Krijgslaan 281 S9, 9000 Gent, Belgium}
\affiliation{$^{36}$ Astronomy Department, Yale University, 52 Hillhouse Ave, New Haven, CT 06511, USA}
\affiliation{$^{37}$ Department of Astronomy, University of Massachusetts, Amherst, MA 01003, USA}

\thanks{$^*$E-mail: \href{mailto:rnaidu@mit.edu}{rnaidu@mit.edu}}
\thanks{$\dagger$ These authors are the PIs of ``Mirage of Miracle" (JWST Program \#5224).}
\thanks{$\ddagger$ NASA Hubble Fellow}

\begin{abstract}
JWST has revealed a stunning population of bright galaxies at surprisingly early epochs, $z>10$, where few such sources were expected. Here we present the most distant example of this class yet -- MoM-z14, a luminous ($M_{\rm{UV}}=-20.2$) source in the COSMOS legacy field at $z_{\rm{spec}}=14.44^{+0.02}_{-0.02}$ that expands the observational frontier to a mere 280 million years after the Big Bang. The redshift is confirmed with NIRSpec/prism spectroscopy through a sharp Lyman-$\alpha$ break and $\approx3\sigma$ detections of five rest-UV emission lines. The number density of bright $z_{\rm{spec}}\approx14-15$ sources implied by our ``Mirage or Miracle" survey spanning $\approx350$ arcmin$^{2}$ is $>100\times$ larger ($182^{+329}_{-105}\times$) than pre-JWST consensus models. The high EWs of UV lines (${\approx}15{-}35$\AA) signal a rising star-formation history, with a ${\approx}10\times$ increase in the last 5 Myr ($\rm{SFR_{\rm{5Myr}}}/\rm{SFR_{\rm{50Myr}}}=9.9^{+3.0}_{-5.8}$). The source is extremely compact (circularized $r_{\rm{e}} = 74^{+15}_{-12}$ pc), and \refrep{yet elongated ($b/a=0.25^{+0.11}_{-0.06}$)}, suggesting an AGN is not the dominant source of \refrep{UV} light. The steep UV slope ($\beta=-2.5^{+0.2}_{-0.2}$) implies negligible dust attenuation and a young stellar population. The absence of a strong damping wing \refrep{provides tentative evidence} that the immediate surroundings of MoM-z14 \refrep{may be} partially ionized at a redshift where virtually every reionization model predicts a $\approx100\%$ neutral fraction. The nitrogen emission and highly super-solar [N/C]$>1$ hint at an abundance pattern similar to local globular clusters that may have once hosted luminous supermassive stars. Since this abundance pattern is also common among the most ancient stars born in the Milky Way, we may be directly witnessing the formation of such stars in dense clusters, connecting galaxy evolution across the entire sweep of cosmic time.
\end{abstract} 


\section{Introduction}
\label{sec:intro}

The quest to observe the earliest galaxies in the universe has been at the heart of observational cosmology since Hubble's discovery of the expanding universe a century ago. JWST was designed with this precise goal in mind: to push the observational frontier to cosmic dawn, revealing the first luminous objects that formed after the Big Bang \citep[e.g.,][]{Gardner06,Gardner23}. Prior to JWST's launch, theoretical models predicted that detecting bright galaxies beyond redshift $z>10$, past the Hubble Space Telescope's reach, would be extraordinarily challenging \citep[e.g.,][]{Mason15,Tacchella18, Williams18}. For example, the sum total of deep and wide surveys planned for the first year of JWST observations \citep[e.g.,][]{Treu22, Eisenstein23, Finkelstein25} were generally expected to yield only a handful of relatively faint ($\gtrsim29-30$ mag AB) sources beyond $z>10$ that would take tens of hours of spectroscopy to confirm \citep[e.g.,][]{Dayal17,Yung19,Behroozi19}. These predictions were built on hierarchical structure formation models, in which early galaxies were expected to be small, faint, and rare \citep[e.g.,][]{White91, BJ05, Springel08}. Broadly speaking, a baseline assumption was that the rate at which gas was transformed into stars (the star-formation efficiency; SFE), the modes of star-formation (e.g., ``bursty" vs. smooth), and the resulting types of stars and black holes at $z>10$ could be predicted based on what was observed at lower redshifts \citep[e.g.,][]{Dave19, Vogelsberger20, Kannan22}.

However, the discovery of GN-z11 at $z\sim11$ with the Hubble Space Telescope \citep[][]{Bouwens10, Oesch16} already provided a first hint that bright galaxies might exist even in the earliest epochs. This remarkable object was both unexpectedly luminous ($M_{\rm{UV}}\approx-21.5$) and detected in a relatively small survey area (within the CANDELS fields spanning a few 100 arcmin$^{2}$; \citealt[][]{Grogin11,Koekemoer11}), suggesting that such galaxies might be more common than predicted. However, without additional examples, it remained unclear whether GN-z11 was an exceptional outlier or representative of a broader population. 

Within weeks of the first science operations, JWST's images revealed an apparent abundance of bright galaxies at photometric redshifts, $z_{\rm{phot}}>10$, challenging pre-JWST consensus models \citep[e.g.,][]{Castellano22,Naidu22,Finkelstein22, Atek23, Harikane23UVLF}. Over two years of photometric surveys continue to report this abundance of luminous sources at the highest redshifts \citep[e.g.,][]{Casey24, Donnan24primer, chemerynska24, Adams24, Kokorev25glimpse, Castellano25, Perez-Gonzalez25}. This unexpected population has electrified the community and raised fundamental questions about galaxy formation in the first $\approx500$ Myrs. However, before we embark on revising the physics of the early Universe, systematic spectroscopic confirmation and characterization of these $z>10$ systems is necessary.

Despite the deluge of candidates reported from photometric surveys, spectroscopic follow-up has been lagging, with only a relatively small number of sources confirmed at $z_{\rm{spec}}>10$ \citep[e.g.,][]{Curtis-Lake23, Bunker23, Wang23UNCOVERz12, Fujimoto23, Harikane24specuvlf, Hsiao24,Napolitano25}. Concerningly, some of the most confidently selected sources have turned out to be low-$z$ interlopers, underscoring the critical need to systematically characterize the contamination fraction in photometric samples \citep[e.g.,][]{Naidu22, Zavala23, Donnan23, ArrabalHaro23, Harikane2025UVLF}. At the bright end ($M_{\rm{UV}}<-20$), where GN-z11 and its peers reside, and at even higher redshifts where the challenge to models is the most acute in the face of the rapidly declining halo mass function, only a handful of sources have been confirmed to date \citep[][]{Castellano24, ArrabalHaro23, Carniani24, Kokorev25}. The most distant among these sources, JADES-GS-z14-0 ($z_{\rm{spec}}=14.18$; \citealt[][]{Carniani24,Carniani25ALMA,Schouws24ALMA}) was found in a mere 10 arcmin$^{2}$ survey implying a $z=14-15$ number density $>100\times$ higher than the pre-JWST theoretical consensus \citep[e.g.,][]{Tacchella18, Robertson24, Whitler25}. Would this remarkable number density hold up if wider areas were systematically surveyed?

The first spectra of luminous $z>10$ sources paint a stunningly rich portrait of the physics of these galaxies. Along many axes, there are few true analogs to these sources in the entire extragalactic Universe. For example, GNz11 \citep[e.g.,][]{Bunker23, Maiolino24GNz11} and GLASS-z12/GHz2 \citep[][]{Naidu22,Castellano22,Castellano24, Zavala25MIRI, Calabro24} display a remarkable chemical abundance pattern (e.g., super-solar [N/O]; e.g., \citealt[][]{Cameron23gnz11}) that prior to JWST had been observed only in a handful of sources \citep[][]{Fosbury03, Patricio16,Mingozzi22}. Might we be witnessing globular cluster formation in action \citep[e.g.,][]{Senchyna24}? Are these compact sources the highest redshift supermassive black holes \citep[e.g.,][]{Maiolino24GNz11}? Are there exotic stellar populations such as supermassive
stars (SMS; $\gtrsim10^{3-4} M_{\rm{\odot}}$) at play \citep[e.g.,][]{Charbonnel23}? Large spectroscopic samples are needed to decide whether the extraordinary chemistry of these sources is ordinary (e.g., generic massive cluster formation; \citealt[][]{Belokurov23}). 

To address the need for spectroscopy at the cosmic frontier, we designed the ``Mirage or Miracle" (MoM) JWST NIRSpec survey (GO-5224, PIs: Oesch \& Naidu; Oesch et al., in prep.). We seek to test whether the extraordinary abundance of bright galaxies at the highest redshifts is a photometric mirage or a spectroscopic miracle, and whether their extraordinary physical conditions are peculiar or commonplace. MoM has systematically targeted a homogeneously selected sample of luminous $z_{\rm{phot}}>10$ galaxies across JWST's legacy wide-area fields (COSMOS, UDS) that were imaged with NIRCam+MIRI by an array of Cycle 1 and 2 programs, most prominently PRIMER \citep[][]{Donnan24primer} and COSMOS-Web \citep[][]{Casey23} (see \S\ref{sec:data} for a full list of imaging programs in these fields utilized in the MoM target selection). 

In this paper we present first results on our primary objective (see \citealt[][]{Naidu25BHstar} for a $z<10$ target). In particular, we present MoM-z14, a luminous ($M_{UV}=-20.2$) galaxy spectroscopically confirmed at $z_\mathrm{spec} = 14.44\pm0.02$. The spectrum not only shows a continuum break from Ly$\alpha$ absorption, but also features UV emission lines, enabling a reliable redshift measurement as well as preliminary constraints on its physical properties. This redshift makes it the most distant spectroscopically confirmed source to date, extending the observational frontier to a mere 280 million years after the Big Bang. The overall survey design and yield, including fillers, as well as the fate of the other targeted $z\gtrsim10$ sources will be presented in subsequent papers.

Throughout this work, we adopt a flat $\Lambda$CDM cosmology with parameters as per \citet[][]{Planck18}. We reference $L^{*}$, the characteristic UV luminosity in Schechter function parametrizations of luminosity functions as per \citet[][]{Bouwens21} which corresponds to $M_{\rm{UV}}\approx-21$ at $z\approx10$. Magnitudes are in the AB system \citep[e.g.,][]{Oke83}. For summary statistics, we typically report medians with uncertainties on the median from bootstrapping (16$^{\rm{th}}$ and 84$^{\rm{th}}$ percentiles). We adopt the following values for solar abundances: $\log(\rm{N/O)}_{\rm{\odot}}=-0.86$, $\log(\rm{C/O)}_{\rm{\odot}}=-0.26$, 12+$\log(\rm{O/H)}_{\rm{\odot}}=8.71$.

\begin{figure*}
    \centering
    \includegraphics[width=18cm]{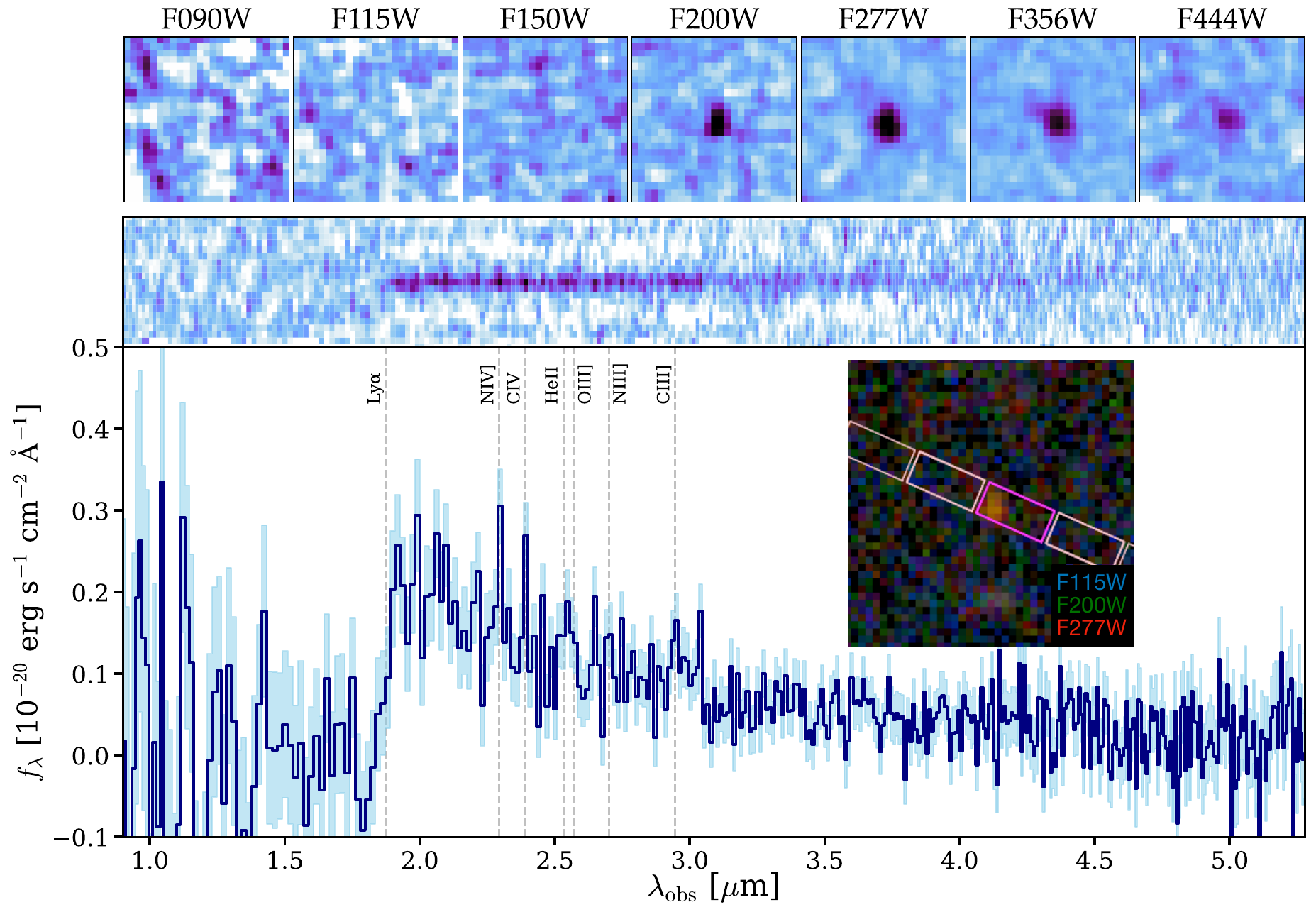}
    \caption{\textbf{JWST imaging and spectroscopy of MoM-z14.} \textbf{Top:} $1\times1''$ NIRCam images spanning 0.9-5$\mu$m show a compact source detected at $\gtrsim2\mu$m that is entirely absent in bluer bands. \textbf{Inset:} NIRCam RGB image with NIRSpec MSA slitlets overlaid from our ``Mirage or Miracle" survey. The source is well-centered and slit losses are modest such that the recovered spectral flux is fully consistent with the NIRCam imaging (see Appendix). \textbf{Bottom:} The prism spectrum (2D SNR spectrum on top, 1D spectrum with 1$\sigma$ errors on the bottom) reveals that the disappearance of the source below $2\mu$m in the imaging is due to an abrupt break whose sharpness implies it is a Lyman-$\alpha$ break. Furthermore, an array of emission lines (dashed lines) supporting the Lyman-$\alpha$ break interpretation is evident. The most prominent among these lines (\nion{N}{iv}]$\lambda1487$\AA, \nion{C}{iv}$\lambda1548,1551$\AA, \nion{C}{iii}]$\lambda1907,1909$\AA) are typically the strongest lines observed in UV spectra of  luminous $z>10$ galaxies \citep[e.g.,][]{Bunker23, Castellano24,Carniani24}.}
    \label{fig:data}
\end{figure*}

\section{Data}
\label{sec:data}

MoM's primary goal is to investigate the nature of luminous ($M_{\rm{UV}}<-20$) $z>10$ photometric candidates. Each of the five NIRSpec pointings in this program -- two in the UDS field and three in COSMOS -- is designed around such candidates as the primary targets. Every pointing receives 4.4 hrs of integration in the $R\approx100$ prism mode. At this depth, the putative Ly$\alpha$ break in our $M_{\rm{UV}}<-20$ candidates may be distinguished from Balmer breaks as well as low-$z$ emission line galaxies. Further details of the MoM sample, survey design, and full survey yield will be described in Oesch et al., in prep.

\subsection{Imaging, Photometry, and Selection of MoM-z14}
\label{sec:imaging}

To build the parent catalog for MoM targets, we combined all publicly available JWST imaging over the COSMOS field as of January 2025 arising from the following programs: COSMOS-Web (\#1727; \citealt[][]{cweb}), Blue Jay (\#1810; \citealt[][]{Belli24}), PRIMER (\#1837; \citealt[][]{Donnan24primer}), JELS (\#2321; \citealt{Duncan24_JELS}), PANORAMIC (\#2514; \citealt[][]{panoramic}), COSMOS-3D (\#5893; \citealt[][]{Kakiichi24}), and SAPPHIRES (\#6434; \citealt[][]{sapphires}). These images are publicly released on the DAWN JWST Archive (DJA)\footnote{\url{https://dawn-cph.github.io/dja/index.html}} as v7.4 COSMOS mosaics. Details about the imaging data reduction using the \texttt{grizli} software \citep[][]{grizli,grizli2} may be found in \citet{Valentino23}. PSF-matched photometric catalogs based on these images were produced following the procedure outlined in \citet{Weibel24}.  Fluxes for the source of interest are listed in Table \ref{table:photom}. \refrep{These are fluxes derived with an adopted aperture radius of $0.16''$ that are then corrected to total following \citet{Weibel24}.}

We fit photometric redshifts to identify potential $z>10$ targets using \texttt{eazy} \citep[][]{Brammer07,Brammer08} with the \texttt{blue\_sfhz} template set featuring thirteen \texttt{FSPS} templates \citep[][]{FSPS1,FSPS2,FSPS3,FSPS4,python-FSPS} and an empirical template based on the strong emission line source at $z=8.5$ studied in \citet[][]{Carnall23}. We also include a template from \citet[][]{Naidu22Schro} fit to a $z=4.9$ dusty emission line galaxy that virtually all photo-$z$ routines place at $z\approx16$ with very high confidence (see discussion in \citealt[][]{ArrabalHaro23, Donnan23, Harikane24specuvlf, Zavala23}).

\begin{figure*}
    \centering
    \includegraphics[width=\linewidth]{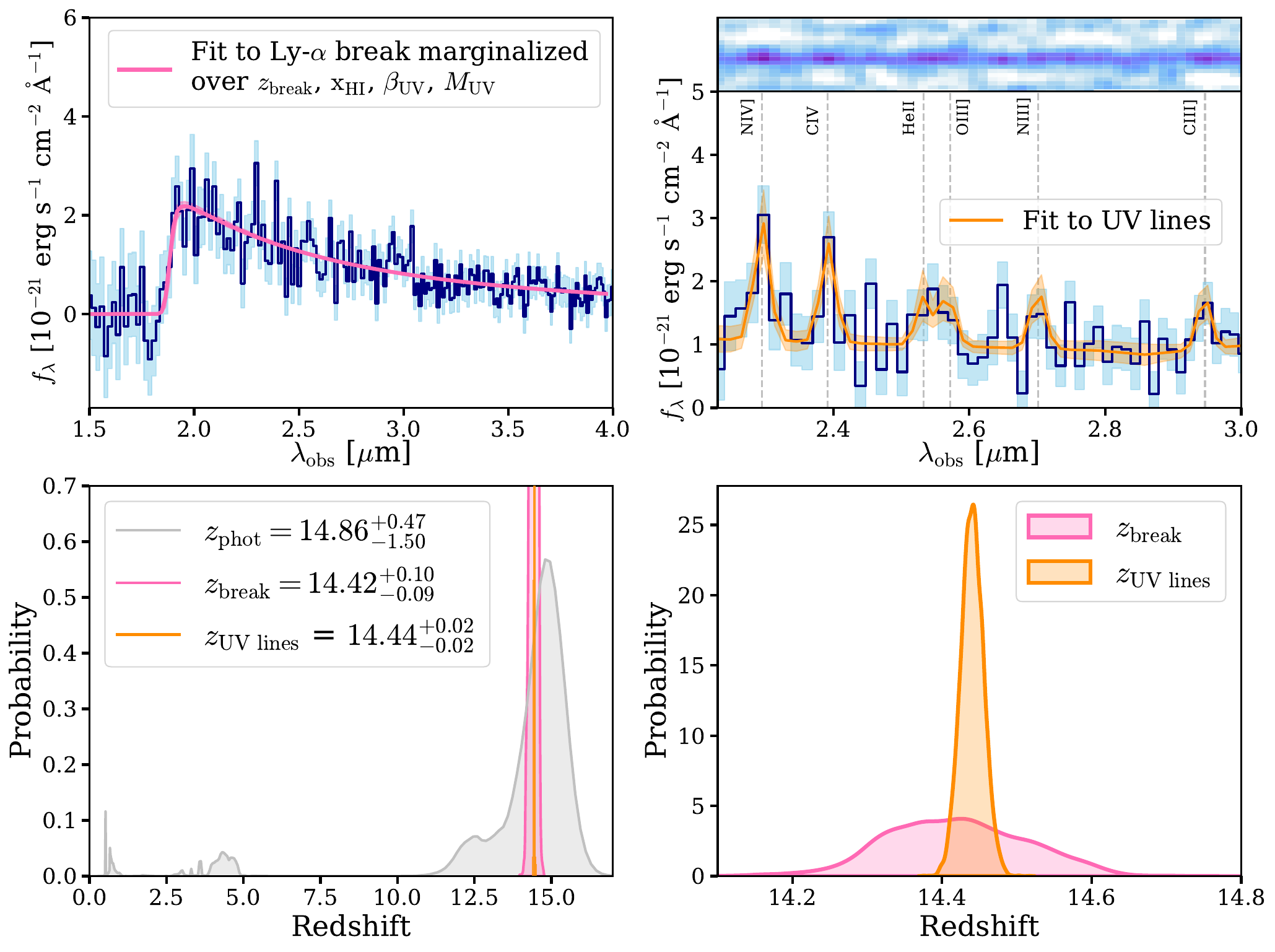}
    \caption{{\bf Summary of spectroscopic redshift constraints.} \textbf{Top-left:} We use the sharp Lyman-$\alpha$ break to derive the redshift ($z_{\rm{break}}$). Our best-fit model (hot pink) accounts for the IGM neutral fraction along the line of sight ($\rm{x}_{\rm{HI}}$) as well as the shape and normalization of the spectrum ($\beta_{\rm{UV}}$, $M_{\rm{UV}}$). See \S\ref{sec:lyaz} for details. \textbf{Top-right:} We are also able to determine the redshift by fitting the rest-UV emission lines that are detected in this source. Each individual line or blend is detected at $\approx3\sigma$. Collectively, this array of lines is detected at $\approx6\sigma$ resulting in an extremely precise redshift. \textbf{Bottom-left:} The photometric redshift distribution derived from NIRCam (silver) shows multiple peaks at $z<5$ (``Schrodinger galaxy"-like solutions; \citealt{Naidu22Schro}) and a dominant $z>10$ solution that led us to target this source. \textbf{Bottom-right:} Comparison of the break redshift and UV line redshift posteriors. Only a handful of galaxies of comparable luminosity at $z>10$ have shown a UV spectrum with multiple lines allowing for a precise redshift determination (\citealt{Bunker23,Castellano24}; both interestingly point-like sources interpreted as putative AGN). Typically, only break redshifts with relatively wide posteriors as shown in pink have been possible \citep[e.g.,][]{Curtis-Lake23, Wang23UNCOVERz12, Carniani24}. }
    \label{fig:specz}
\end{figure*}

We based one of the five MoM pointings around MoM-z14 as a high priority target. It stood out to us as a robust drop out, completely disappearing in the F090W, F115W, and F150W filters (see Fig. \ref{fig:data}). The photometric redshift of this source is $z_{\rm{phot}}=14.86^{+0.47}_{-1.50}$, with a strong preference for $z>10$ and marginal low-$z$ solutions (bottom-left panel, Fig. \ref{fig:specz}). Importantly, this high-$z$ solution also survives the inclusion of the $z=4.9$ interloper template.

\subsection{Spectroscopy}
\label{sec:spec}

\begin{figure*}
    \centering
\includegraphics[width=0.8\linewidth]{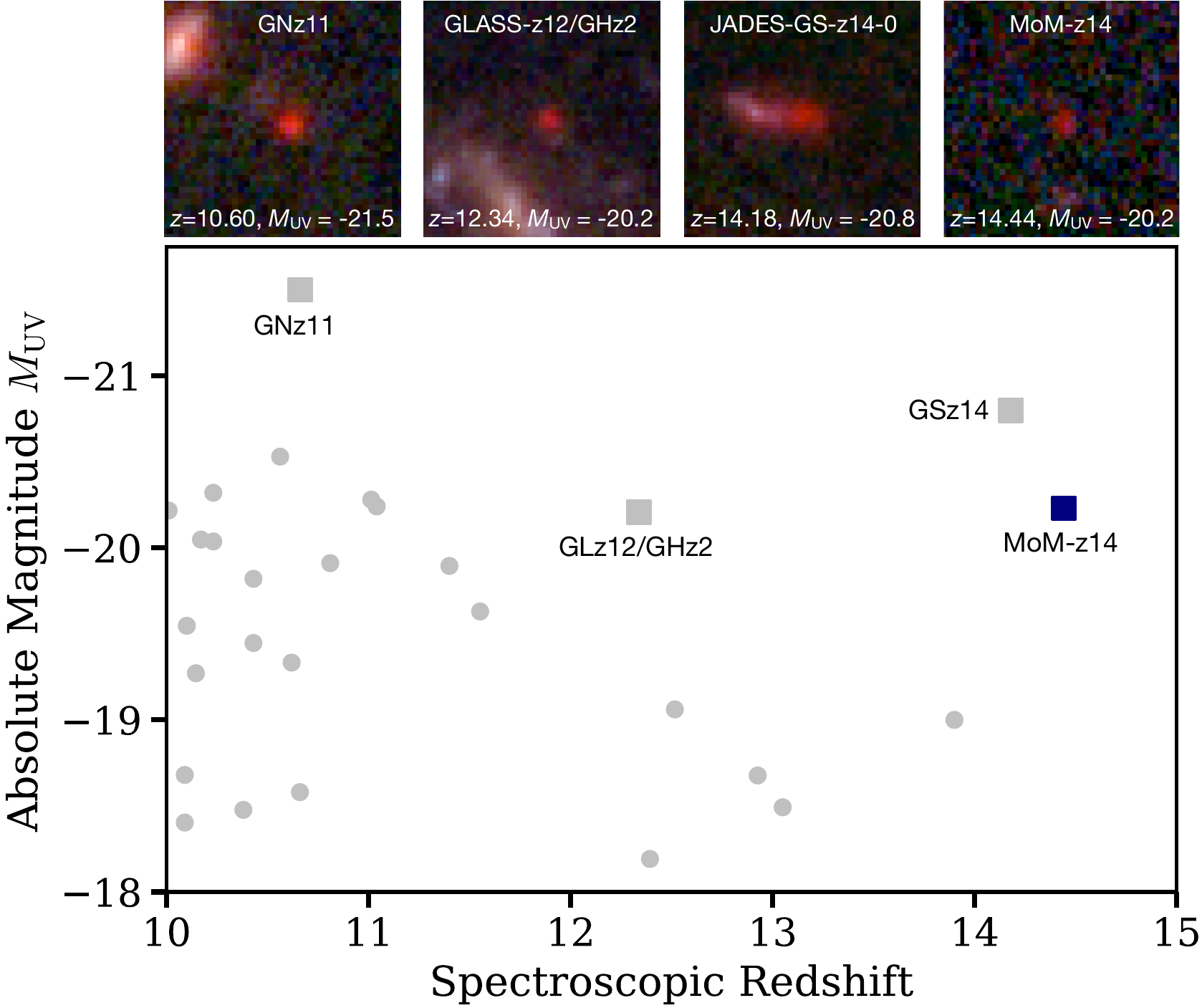}
    \caption{{\bf  Compilation of absolute UV magnitude vs. spectroscopic redshift for sources at the cosmic frontier.} \textbf{Bottom:} The galaxies shown in silver arise from $\approx600$ arcmin$^{2}$ surveyed by JWST in the first $\approx2.5$ years of its operations. See text for a full list of references. We highlight three of the most well-studied bright sources at these epochs that we reference regularly as points of comparison for MoM-z14 with square markers -- GNz11 \citep[][]{Oesch16,Bunker23,Tacchella23GNz11}, GLASS-z12/GHz2 \citep[][]{Naidu22,Castellano22, Castellano24}, and JADES-GS-z14-0 \citep[][]{Robertson24JOF,Carniani24}. \textbf{Top:} $1''\times1''$ RGB stamps (F090W, F115W, F277W) of the sources highlighted with square markers. Three of these sources are extremely compact, with GS-z14 being the exception. For the silver points above, we extend the compilation from \citet{Roberts-Borsani24} and include the following papers: \citet{Curtis-Lake23,  Wang23UNCOVERz12, Fujimoto23,ArrabalHaro23, Harikane24specuvlf, Hsiao24,Napolitano25, Kokorev25, Witstok25}.}
    \label{fig:muvz}
\end{figure*}

MoM-z14 was targeted in one of the COSMOS MoM masks observed on 16th April, 2025. We reduced its spectrum using \texttt{v0.9.4} of the \texttt{msaexp} software \citep[][]{msaexp}, details of which can be found in \citet{Heintz24,degraaff24}. The reduction choices we employ are the exact same as the upcoming v4 NIRSpec release of the DAWN JWST Archive (Pollock et al., in prep.; \citealt{Valentino25}). A key development in this version compared to \citet{Heintz24,degraaff24} is an updated flux calibration and an expansion of the nominal prism wavelength coverage up to $5.5\mu$m. The processed prism spectrum (4.4h exposure time) is shown in Fig. \ref{fig:data} along with the well-centered position of the source in its assigned MSA slitlets. 

We perform two data validation checks that are shown in the Appendix. First, the shape and normalization of the prism spectrum is in excellent agreement with the NIRCam photometry (Fig. \ref{fig:nircamvnirspec}). Note that the photometry is not used in any step of the data reduction, and so this is an independent validation of the overall flux calibration. Next, we reduce the spectrum with a ``global'' sky background where the sky spectrum is estimated from empty slitlets on the mask. This is in contrast to the fiducial spectrum based on the ``local" differencing at different nod positions. Given the high background in COSMOS owing to its proximity to the ecliptic (relative to, e.g., EGS, Abell 2744, GOODS-N), this is a potentially important systematic. Importantly, all the key features analyzed in this work (e.g., the sharp break, UV emission lines) are detected in both versions (see Fig. \ref{fig:globalsky}).

\begin{deluxetable}{lrrrrrrrrrrr}
\tabletypesize{\footnotesize}
\tablecaption{Summary of Results}
\tablehead{\multicolumn{2}{c}{Empirical Properties}
\label{tab:properties}
}
\startdata
\vspace{-0.3cm}\\
R.A. [deg] & 150.0933255\\
Dec. [deg] & 2.2731627\\
Redshift (UV lines) & $14.44^{+0.02}_{-0.02}$\\
Redshift (Ly$\alpha$ break) & $14.42^{+0.10}_{-0.09}$\\
UV Luminosity ($M_{\rm{UV}}$) & $-20.23^{+0.06}_{-0.06}$\\ 
UV slope ($\beta_{\rm{UV}}$; $f_{\lambda}\propto\lambda^{\beta}$) & $-2.47^{+0.17}_{-0.17}$\\
Galaxy size (circularized $r_{\rm{e}}$) [pc] & $74^{+15}_{-12}$\\
Galaxy size (semi-major axis $a$) [pc] & $147^{+19}_{-20}$\\
Axis ratio ($b/a$) & $0.25^{+0.11}_{-0.06}$\\
\hline\vspace{-0.2cm}\\
\multicolumn{2}{c}{\texttt{Prospector} SED modeling}\vspace{0.05cm} \\
\hline \vspace{-0.15cm}\\
Stellar Mass (log($M_{\rm{*}}/\rm{M}_{\rm{\odot}}$)) & $8.1^{+0.3}_{-0.2}$\\
Star-Formation Rate (${\rm{5\ Myr}}$ )[$\rm{M}_{\rm{\odot}}$ yr$^{-1}$] & $13.0^{+3.7}_{-3.5}$\\
Star-Formation Rate (${\rm{50\ Myr}}$) [$\rm{M}_{\rm{\odot}}$ yr$^{-1}$] & $2.2^{+1.5}_{-0.6}$\\
Dust Attenuation ($A_{\rm{5500\AA}}$) & $0.2^{+0.2}_{-0.1}$ \\
Age ($t_{\rm{50}}$/Myr) & $4.0^{+10.0}_{-1.4}$\\
Star-Formation Surface Density [$M_{\rm{\odot}}\ \rm{yr}^{-1}\ \rm{kpc}^{-2}$] & $233^{+107}_{-107}$\\ 
Stellar Surface Density ($\log(\Sigma_{\rm{*}}/M_{\rm{\odot}}\ \rm{kpc}^{-2}$)) & $9.6^{+0.2}_{-0.7}$\\
\hline\vspace{-0.2cm}\\
\multicolumn{2}{c}{\texttt{Cue} Emission Line Modeling}\vspace{0.05cm} \\
\hline \vspace{-0.15cm}\\
Ionization Parameter ($\log U$) & $-1.54^{+0.39}_{-0.54}$ \\
Ionizing Efficiency ($\log (\xi_\mathrm{ion}/\rm{erg^{-1}\ s^{-1}})$) & $26.28^{+0.45}_{-0.49}$\\
Gas Density ($\log (n_\mathrm{H}/\rm{cm}^{-3})$) & $3.03^{+0.66}_{-0.96}$ \\
Oxygen abundance $\rm{[O/H]}$ & $-1.38^{+0.65}_{-0.56}$ \\
Nitrogen-to-Oxygen ratio $\rm{[N/O]}$ & $0.29^{+0.28}_{-0.45}$ \\
Carbon-to-Oxygen ratio $\rm{[C/O]}$ & $-0.65^{+0.39}_{-0.22}$\\
Nitrogen-to-Carbon ratio $\rm{[N/C]}$ & $0.90^{+0.29}_{-0.63}$\\
Nitrogen-to-Carbon ratio (from line ratios, $\rm{[N/C]}$)$^{\rm{a}}$ & $>1$ (2$\sigma$)
\enddata
\tablecomments{$^{\rm{a}}$Following \citet{Villar-Martin04} who analyzed the N-emitting Lynx arc at $z=3.36$, we conservatively bracket $T_{\rm{e}}/(10^{4}\rm{K})=0.5-3$ \citep[e.g.,][]{Cameron23gnz11,Calabro24,Carniani25CII} and derive ionic abundances from \nion{C}{iv}, \nion{C}{iii}],\nion{N}{iv}], and \nion{N}{iii}].}
\end{deluxetable}

\begin{deluxetable}{lc}
\tablewidth{0.4\linewidth}
\centering
\tabletypesize{\footnotesize}
\tablecaption{PSF-matched photometry \label{table:photom}}
\tablehead{
\colhead{Band} & \colhead{Flux [nJy]}
}
\multicolumn{2}{c}{HST (ACS, WFC3)} \\
\hline \vspace{-0.18cm}\\
F606W & $6\pm6$ \\
F814W & $-6\pm6$ \\
F125W & $15\pm15$ \\
F140W & $1\pm15$ \\
F160W & $10\pm14$ \\
\hline\vspace{-0.18cm} \\
\multicolumn{2}{c}{JWST (NIRCam)} \\
\hline\vspace{-0.18cm} \\
F090W & $0\pm6$ \\
F115W & $-1\pm5$ \\
F150W & $4\pm4$ \\
F200W & $14\pm3$ \\
F277W & $22\pm2$ \\
F356W & $20\pm2$ \\
F410M & $13\pm4$ \\
F444W & $13\pm3$
\enddata
\tablecomments{\refrep{Total} fluxes derived following \citet[][]{Weibel24} based on \texttt{v7.4} COSMOS mosaics hosted on the DAWN JWST Archive. The aperture radius adopted is $0.16''$.}
\end{deluxetable}

\begin{deluxetable}{lrrrrrrrrrrr}
\centering
\tabletypesize{\footnotesize}
\tablecaption{Emission Line Measurements}
\tablehead{
\colhead{Line} & \colhead{Flux} & \colhead{EW$_{\rm{0}}$} & \colhead{SNR}
}
\startdata
\label{tab:lines}
\vspace{-0.3cm}\\
\nion{N}{iv}]$\lambda1487$\AA & $54.3^{+16.1}_{-15.5}$ & $33.3^{+15.4}_{-11.8}$ & 3.4 \\
\nion{C}{iv}$\lambda1548,1551$\AA & $44.0^{+12.8}_{-12.8}$ & $27.6^{+11.2}_{-9.2}$ & $3.4$\\
\nion{He}{ii}$\lambda1640$\AA+\nion{O}{iii}]$\lambda1661,1666$\AA & $45.3^{+15.6}_{-14.8}$ & $29.9^{+12.3}_{-10.5}$ & $3.0$\\
\nion{N}{iii}]$\lambda1747-1754\rm{\AA}^{\rm{a}}$ & $25.2^{+9.9}_{-8.9}$ & $17.4^{+8.6}_{-6.7}$ & 2.7\\
\nion{C}{iii}]$\lambda1907,1909$\AA & $21.5^{+7.7}_{-7.6}$ & $15.0^{+6.7}_{-5.7}$ & $2.8$
\enddata
\tablecomments{Fluxes are in units of $10^{-20}$ erg s$^{-1}$ cm$^{-2}$ and EWs in \AA\ (rest-frame). FWHM for all lines is allowed to range up to 650 km s$^{-1}$, with the width of all lines tied to each other -- the FWHM is completely unconstrained in our fits. $^{\rm{a}}$Quintuplet comprising lines at 1747\AA, 1749\AA, 1750\AA, 1752\AA, 1754\AA.}
\end{deluxetable}

\section{Results}
\label{sec:data}

\subsection{Spectroscopic Redshift}
\label{sec:specz}

\subsubsection{UV lines}

We use \texttt{UNITE}, a custom NIRSpec emission line fitting package \citep[][]{Hviding25}, to model the UV emission lines. This package self-consistently handles the translation between idealized models of the spectrum to realistic observed NIRSpec spectra by accounting for e.g., the wavelength-dependent resolution, LSF (idealized for a point-source, applicable to our relatively compact object) and calibration uncertainties following \citet{degraaff24jades,degraaff24rubies}. We initialize Gaussians to represent the typical UV lines seen in spectra of luminous $z>10$ galaxies \citep[e.g.,][]{Bunker23,Maiolino24GNz11,Castellano24} with uniform priors on their FWHM up to 650 km s$^{-1}$ and redshift between $z=14-15$. The width and redshift are tied across all lines. \refrep{By construction, with \texttt{UNITE} the fitted widths of the emission lines are consistent with the NIRSpec LSF.}  Various multiplets and adjacent lines are blended at the prism resolution -- we model these by splitting the lines \refrep{in a multi-Gaussian fit} but report combined fluxes by summing their posterior fluxes. The continuum is fit by considering a region of $\pm15,000$ km s$^{-1}$ around each line, and masking a central region of $\pm3,500$ km s$^{-1}$. Overlapping regions around adjacent emission lines are stitched together and treated as a contiguous region. The posteriors are sampled using \texttt{numpyro} \citep[][]{numpyro} and are shown in Fig. \ref{fig:specz} and reported in Table \ref{tab:lines}. 

We recover five UV lines at $\approx3\sigma$ (top-right panel, Fig. \ref{fig:specz}). \refrep{There are also hints of additional lines} (e.g., \nion{Ne}{iv} and \nion{C}{ii}) \refrep{albeit at lower significance ($<2\sigma$) that will require deeper data to confirm}. Reassuringly, the reported lines correspond to peaks in the 2D spectrum, and are also seen in a version of the spectrum processed using the global sky background (Fig. \ref{fig:globalsky}). The strongest lines detected -- forbidden and semi-forbidden transitions of nitrogen and Carbon -- also happen to be the strongest UV lines seen in other comparably luminous $z>10$ galaxies (e.g., \citealt[][]{Bunker23,Castellano24,Carniani24}). 

Owing to these UV lines, the inferred redshift is quite precise ($z=14.44^{+0.02}_{-0.02}$), which is unique for a prism redshift at this early epoch \citep[e.g.,][]{Curtis-Lake23, Wang23UNCOVERz12,Carniani24}. This will enable extremely efficient ALMA spectral scan follow-up for e.g., [\nion{O}{iii}]88$\mu$m and [\nion{C}{ii}]158$\mu$m as well as precise modeling of the damping wing. Typically, due to the paucity of detected lines at prism resolution, only Ly$\alpha$ break redshifts with much wider posteriors have been possible (see bottom panels of Fig. \ref{fig:specz}). This is due to degeneracies between the location of the break, the neutral fraction of the IGM, and the presence of damped Ly$\alpha$ absorption -- e.g., $z_{\rm{prism}}=14.32^{+0.08}_{-0.20}$ for JADES-GS-z14-0 subsequently refined with ALMA to $z_{\rm{[OIII]}}=14.1793^{+0.0007}_{-0.0007}$ \citep[][]{Carniani24,Schouws24ALMA,Carniani25ALMA}.

\subsubsection{Lyman-$\alpha$ break} \label{sec:lyaz}

We perform an independent verification of the emission-line redshift by modeling the observed break as a Ly$\alpha$ break at a redshift of $z_{\rm{Ly\alpha}}=14-15$. The sharpness of the break leaves little doubt that this is indeed a Ly$\alpha$ break -- see Fig. \ref{fig:lowz} in the Appendix for a comparison against a typical Balmer break spectrum that provides a poor match to the shape (a smooth, gradual decline vs. the sharp edge) and depth of the break (flux would be detected blueward of the break at the depth of these observations). \refrep{A recently discovered class of objects -- so called ``Black Hole Stars" \citep[e.g.,][]{Naidu25BHstar,degraaff25} thought to power the Little Red Dots \citep[e.g.,][]{Matthee24} -- show the largest Balmer breaks on record that may mimic Lyman breaks in broadband photometry, but these sources are excluded based on their characteristic red optical continuum and broad Balmer lines that are not seen in the spectrum.}

Here we describe our model set-up to fit the Ly$\alpha$ break redshift. The prescription from \cite{Miralda-Escude98} is used to approximate the absorption profile due to the Gunn-Peterson effect \citep[][]{Gunn65} from neutral hydrogen (H\,{\sc i}) in the intergalactic medium (IGM). The correction to this profile from \citet{Totani06} is used to determine the average neutral H\,{\sc i} fraction, $x_{\rm HI}$, of the IGM in the line of sight up to the Ly$\alpha$ redshift, $z_{\rm Ly\alpha}$. We assume an underlying intrinsic galaxy continuum described by a power-law, $F\propto F_{\rm 0}\lambda^{\beta_{\rm UV}}$, where $\beta_{\rm UV}$ is the rest-frame UV spectral slope and $F_0$ is an arbitrary normalization factor. We use {\tt dynesty} \citep{Speagle19} to estimate the posteriors on the free parameters in the model ($F_0$, $\beta_{\rm UV}$, $x_{\rm HI}$, and $z_{\rm Ly\alpha}$) and the total evidence of the distribution. We convolve the model to match the wavelength-dependent spectral resolution of the data. \refrep{Emission lines are fit simultaneously along with the continuum slope.} 

We find a best-fit Ly$\alpha$ break redshift of $z_{\rm Ly\alpha} = 14.42^{+0.10}_{-0.09}$, consistent with the photometric redshift and the inferred emission-line redshift (see Fig.~\ref{fig:specz}). Intriguingly, the best-fit IGM neutral fraction is non-unity at an epoch where virtually all reionization models predict $\rm{x}_{\rm{HI}}\approx100\%$ \citep[e.g.,][]{Finkelstein19, Mason19nonpar,Naidu20,Hutter21a, MN22, Kannan22}. We also do not require a strong DLA to model this break, which is uncommon at these redshifts \citep[e.g.,][]{Curtis-Lake23,Carniani24,Heintz24_DLA}. We will discuss these constraints in detail in \S~\ref{sec:Lya}. 

\subsection{Physical Properties}

\subsubsection{Morphology}
\label{sec:morphology}

\begin{figure}
    \centering
    \includegraphics[width=\linewidth]{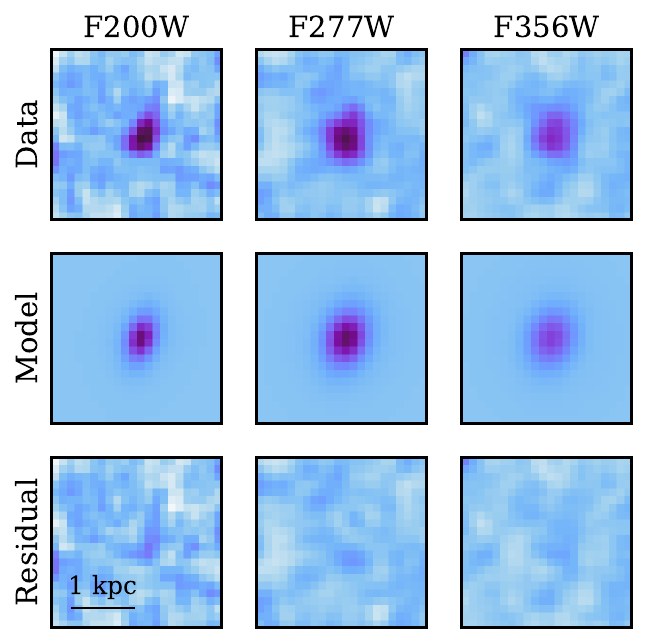}
    \caption{{\bf Results of morphology analysis.} The model built with \texttt{forcepho} (middle row) is fit to the three bands simultaneously and provides a satisfactory match to the data (top row) with no appreciable residuals (bottom row). An extremely compact (circularized $r_{\rm{50}} = 74^{+15}_{-12}$ pc) source is recovered in these fits. From these fits it is clear that the object is not dominated by a point-source, as may be expected if its UV luminosity was dominated by an AGN.}
    \label{fig:morphology}
\end{figure}

\begin{figure*}
    \centering
    \includegraphics[width=\linewidth]{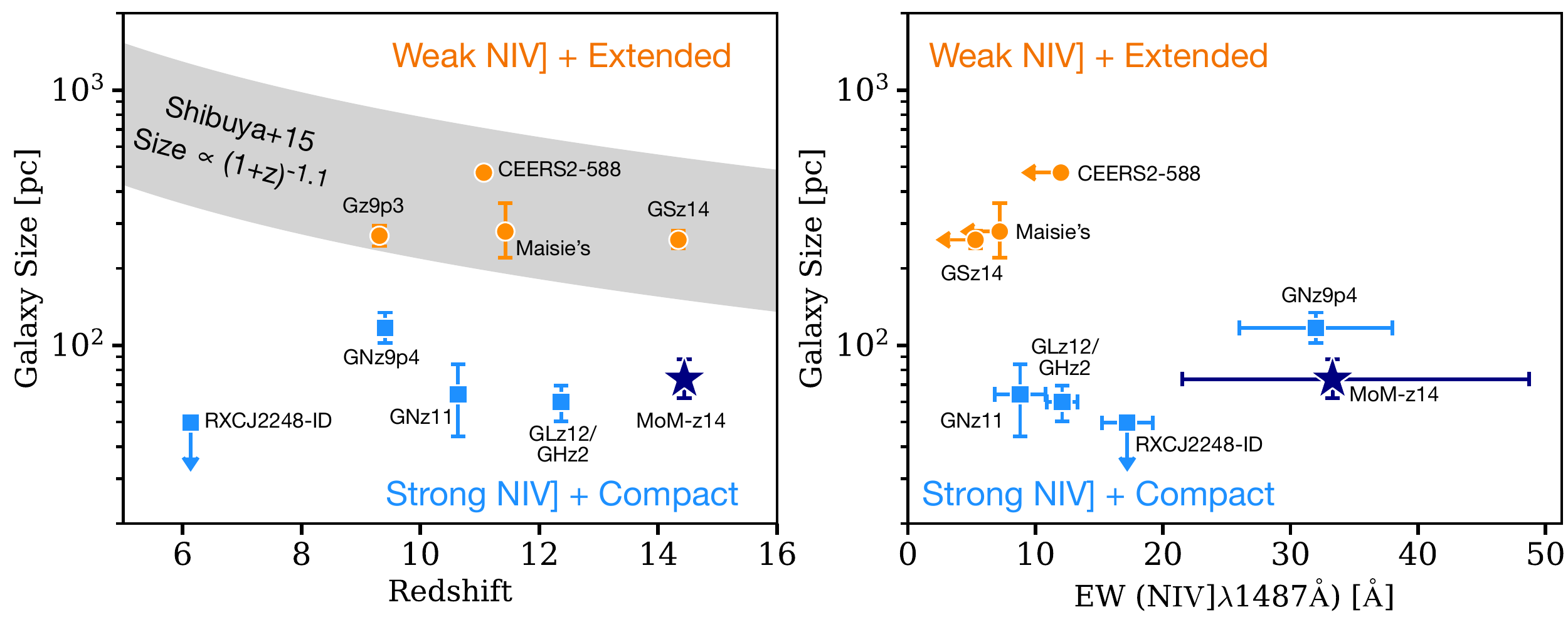}
    \caption{{\bf A size-abundance bimodality among bright $z>10$ galaxies.} \label{fig:bimodal}Figure adapted from \citet{Harikane2025UVLF} who reported this trend (see also \citealt[][]{Schaerer24}). Two types of sources are evident in the size-redshift plane (left) and size-EW plane (right) -- compact \nion{N}{iv}] emitters such as MoM-z14 (purple, navy) and extended sources with weak \nion{N}{iv}] such as JADES-GS-z14-0 (orange). The extrapolated size evolution from \citet{Shibuya15} (silver band) is shown in the left panel for guidance. MoM-z14 joins GNz11 and GLASSz12/GHz2 as extremely compact outliers at $z>10$, with sizes $<5-10\times$ what would be expected from the scaling relation. The measurements shown here are compiled from the following sources: GNz11 \citep[][]{Tacchella23GNz11, Bunker23}, GLASS-z12/GHz2 \citep[][]{Castellano24}, GNz9p4 \citep[][]{Schaerer24}, JADES-GS-z14-0 \citep[][]{Carniani24}, Maisie's Galaxy \citep[][]{Finkelstein23, ArrabalHaro23}, RXCJ2248-ID \citep[][]{Topping24}, Gz9p3 \citep[][]{Boyett24}, and CEERS2-588 \citep[][]{Harikane24specuvlf}. }
\end{figure*}

The morphology of MoM-z14 is of key interest to understand the origins of its UV luminosity. Among the handful of spectroscopically confirmed bright $z>10$ objects, there appear to be two types -- point-sources indistinguishable from the PSF (e.g., GNz11 and GLASS-z12/GHz2), and extended galaxies (e.g., JADES-GS-z14-0 and ``Maisie's Galaxy"; \citealt{Finkelstein22Maisies}) with sizes following empirical scaling relations from lower redshifts (e.g., $r_{\rm{50}}\approx(1+z)^{-1}$; \citealt[][]{Shibuya15}). Furthermore, as noticed by \citet{Harikane24}, these morphological differences are reflected in chemical abundance patterns, signaling a deeper connection between morphology and evolutionary pathways (Fig. \ref{fig:bimodal}, and further discussed in \S\ref{sec:gcs}).

We model MoM-z14 using the \texttt{forcepho} software (Johnson et al., in prep.; recently deployed in e.g., \citealt[][]{Tacchella23GNz11,Tacchella23SMACS,Robertson23,Carniani24}). Briefly, \texttt{forcepho} is a Bayesian modeling framework that simultaneously fits PSF-convolved Sersic profiles to all filters under consideration (in our case F200W, F277W, and F356W where the source is well-detected). The PSFs (from WebbPSF; \citealt{Perrin14}) are approximated as Gaussian mixture models for rapid convolution. In addition to using Hamiltonian Monte Carlo to sample from the posterior, a key feature of \texttt{forcepho} is that it is run on individual processed exposures and not mosaicked images, thereby circumventing issues such as, e.g., distortions of the PSF while also making maximal use of sub-pixel information that is smeared out in mosaics.

We present the \texttt{forcepho} model for MoM-z14 in Fig. \ref{fig:morphology}. The thumbnails shown are 30 mas/pixel images corresponding to 98 pc at $z=14.44$. We find the source is resolved and compact (circularized $74^{+15}_{-12}$ pc; semi-major axis, $a=147^{+19}_{-20}$ pc) with a Sersic index of $1.0^{+0.2}_{-0.2}$. The source is extended along the North-South direction in the thumbnails shown, with an elongated axis ratio ($b/a=0.25^{+0.11}_{-0.06}$). \refrep{Further evidence for this elongated morphology is seen in the radial surface brightness profiles that are extended relative to the PSF (see Fig. \ref{fig:radial_profiles} in Appendix)}. There may be a hint of a second component in the images responsible for this elongation. We defer detailed investigation of a two-component model with deeper imaging to future work. We have verified that we find similar results for the size ($\approx90$ pc) by modeling the source with \texttt{pysersic} \citep[][]{Pasha23} directly on the mosaicked images using empirical PSFs constructed following \citet{Weibel24}. 

MoM-z14 deviates from typical size-luminosity-redshift scaling relations, joining GLASS-z12/GHz2 and GNz11 as outliers at $z>10$ in terms of being extremely compact despite showing high UV luminosity (Fig. \ref{fig:bimodal} shows the scaling for $M_{\rm{UV}}\sim-21$ mag sources; \citealt[][]{Shibuya15}). Further, as we will discuss in \S\ref{sec:gcs}, MoM-z14 shows similar indications of super-solar nitrogen suggesting a common evolutionary channel for these compact sources. However, it is worth noting that, while compact, MoM-z14 is not dominated by a central point-source unlike these objects that disfavor an AGN as the dominant source of the UV light \citep[][]{Tacchella23GNz11, Ono23, Maiolino24GNz11}.

\subsubsection{SED Modeling}
\label{sec:prosp}

\begin{figure*}
    \centering
    \includegraphics[width=\linewidth]{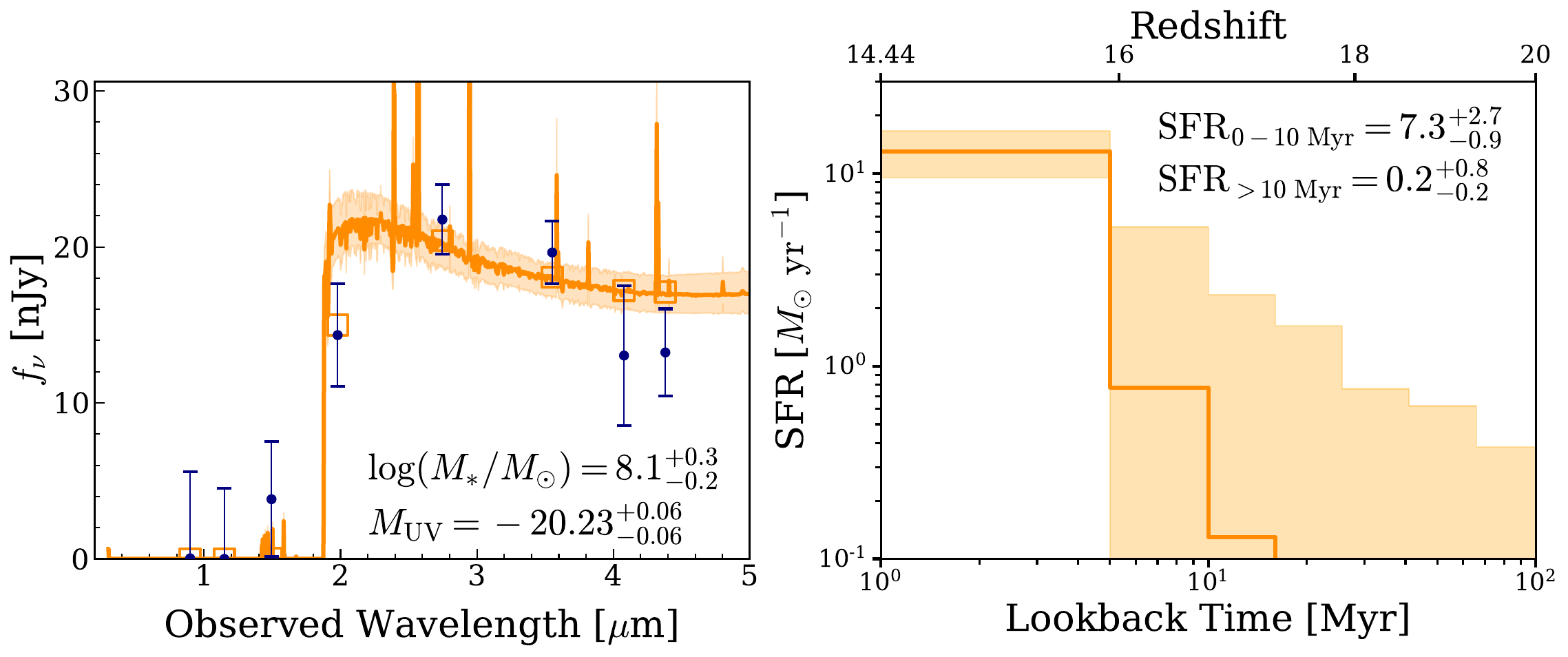}
    \caption{\textbf{SED fitting results using \texttt{Prospector}} \citep[][]{Johnson21}. \textbf{Left:} The model (orange) is fit to the NIRCam photometry (navy) and \nion{C}{iii}]$\lambda1907,1909$\AA\ emission line flux. Modeling all the observed emission lines simultaneously requires a more flexible model that we explore in \S\ref{sec:cue} and Fig. \ref{fig:cue}. \textbf{Right:} The inferred star-formation history shows a steep rise, with as much as a $>10\times$ increase over the last $\approx10$ Myrs. The high \nion{C}{iii}]$\lambda1907,1909$\AA\ EW that can be matched only with a young ($<5$ Myr) burst \citep[e.g.,][]{Jaskot16} informs this rising SFH. Note, however, that an additional older stellar population cannot be ruled out with these data as evidenced by the broad posteriors.}
    \label{fig:prosp}
\end{figure*}

We model the SED using the \texttt{prospector} Bayesian modeling framework \citep{Leja17, Leja19, Johnson21}. Our setup closely follows the choices validated in \citet[][]{Tacchella22,Naidu22,NM24}. We use FSPS \citep{FSPS1,FSPS2,FSPS3} with the \texttt{BPASS} stellar models \citep[][]{BPASS}, in particular the \texttt{v2.2 -bin-imf135all 100} models that assume a \citet{Salpeter55} IMF with a $100 M_{\rm{\odot}}$ cutoff. Nebular emission is modeled with the \texttt{CLOUDY} \citep{Ferland17} grid produced in \citet{Byler17}. The parameters we fit include seven bins describing a non-parametric star-formation history, the total stellar mass, stellar and gas-phase metallicities, nebular emission parameters, and a flexible dust model \citep{KriekConroy13}. We adopt a ``bursty continuity" prior for the star-formation history \citep{Tacchella22} with the time bins logarithmically spaced up to a formation redshift of $z=20$. We hold the first two bins fixed at lookback times of 0-5 and 5-10 Myr to capture bursts that power strong emission lines \citep[e.g.,][]{Whitler22,Tacchella23SMACS} that are increasingly ubiquitous towards the Epoch of Reionization and beyond \citep[e.g.,][]{Matthee23,Meyer24,CoveloPaz25, Lin25CONGRESS}. Posteriors are sampled using \texttt{dynesty} \citep[][]{dynesty}. It is important to note that systematic uncertainties loom over our model assumptions. For example, it is unclear whether our adopted IMF is applicable to a luminous $z=14.44$ galaxy \citep[e.g.,][]{Cameron24,Hutter25, Yung24}. The SED fitting results must therefore be viewed as a baseline set of quantities derived under canonical assumptions. 

We fit the model at fixed redshift to the NIRCam photometry and the integrated \nion{C}{iii}] emission line flux. Of all the lines observed, \nion{C}{iii}] is modeled best by the \citet{Byler17} grid. The other lines are interpreted in the following section using a more flexible model (\texttt{Cue}; \citealt[][]{Li24cue}) that is capable of capturing, e.g., super-solar [N/O] to produce strong nitrogen lines. The posteriors of the SED and star-formation history (SFH) are shown in Fig. \ref{fig:prosp}. The quality of the fit is reasonable ($\chi^{2}/N = 1.2$), with the key area for improvement being the \nion{C}{iii}] line flux that is underestimated by 0.3 dex. We note that this disagreement is $2\times$ worse when using the \texttt{MIST} \citep[][]{Choi17, Choi20} stellar library with a \citet{Chabrier03} IMF. 

We infer MoM-z14 to be a relatively low-mass galaxy, comparable to the present-day Small Magellanic Cloud ($\approx10^{8} M_{\rm{\odot}}$; \citealt[][]{vandermarel09}). There is negligible dust attenuation, as evidenced by the blue UV slope ($\beta_{\rm{UV}}\approx-2.5$). We are observing this source during a burst phase, wherein the star-formation rate may have increased by up to an order of magnitude in a short span of $\approx10$ Myr. We have verified that the high \nion{C}{iii}] EW is driving this rise in the SFH -- a fit without this line returns a prior-dominated, relatively flat SFH. Indeed, dedicated grids built to explore \nion{C}{iii}] show that $\approx15$\AA\ EWs signal $\lesssim5$ Myr bursts, high ionization parameters ($\log U\gtrsim-2$), and low gas-phase metallicities ($\lesssim10\% Z/Z_{\rm{\odot}}$) \citep[e.g.,][]{Jaskot16, Nakajima18}. While the galaxy is in the throes of a burst, we cannot rule out the presence of even older stellar populations as reflected in the SFH posteriors, given that we are working purely with the rest-UV. Detailed chemical abundances from high-resolution spectroscopy (e.g., with NIRSpec, ALMA) may be the most efficient path to refining the SFH as the relative contributions of ``delayed" channels (e.g., neutron star mergers, low-mass stars) and ``prompt" channels (e.g., supermassive stars) manifest in different patterns \citep[e.g.,][]{Kobayashi20,Johnson23, Kobayashi24}. 

\subsubsection{Emission Line Modeling}
\label{sec:cue}

\begin{figure*}
    \centering
\includegraphics[width=\linewidth]{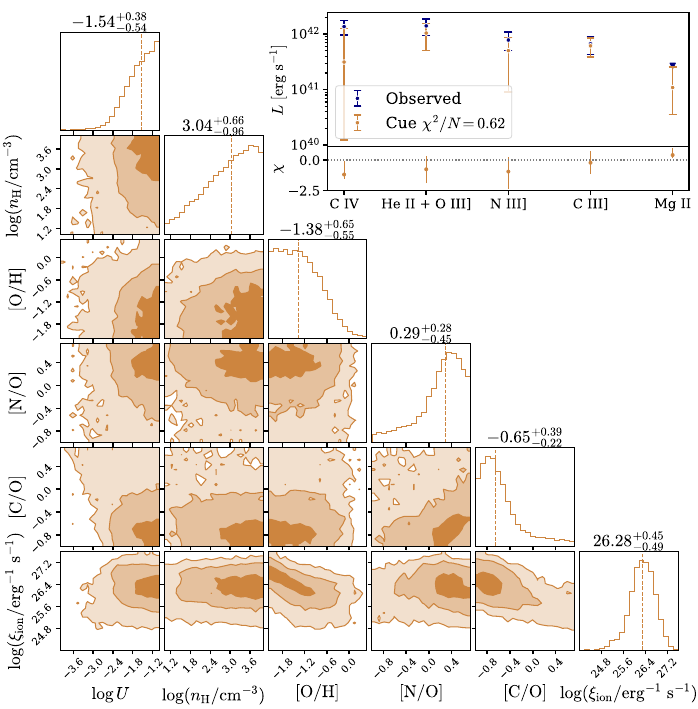}
    \caption{\textbf{Emission line modeling results using \texttt{Cue}} \citep[][]{Li24cue}. MoM-z14's UV spectrum, with hints of several emission lines, provides a unique window into the physics of bright galaxies at cosmic dawn -- the ionizing sources powering them ($\log \xi_{\rm{ion}}$), the state of the gas in their ISM ($\log n_{\rm{H}}$, $\log U$), and their chemical abundance patterns ([O/H], [N/O], [C/O]). Satisfactory fits to the emission lines are shown in the top-right panel, while the corner-plot illustrates the inferred posteriors. Even with only SNR$\approx3$ emission lines, and despite the low-resolution of the data (e.g., \nion{He}{ii}$\lambda1640$\AA+\nion{O}{iii}]$\lambda1661,1666$\AA\ are observed as a blend), the model is not entirely unconstrained. Consistent with the lack of neutral gas around the source, a highly ionizing radiation field is inferred (e.g., $\log \xi_{\rm{ion}}/\rm{erg\ s^{-1}}\approx26.3$). There are also indications of a super-solar [N/O] and sub-solar [C/O] abundance pattern already in place by $z=14.44$ reminiscent of GN-z11 \citep[e.g.,][]{Cameron23} and other recently discovered N-emitters \citep[e.g.,][]{Schaerer24}. 
    } 
    \label{fig:cue}
\end{figure*}

\begin{figure}
    \centering
\includegraphics[width=\linewidth]{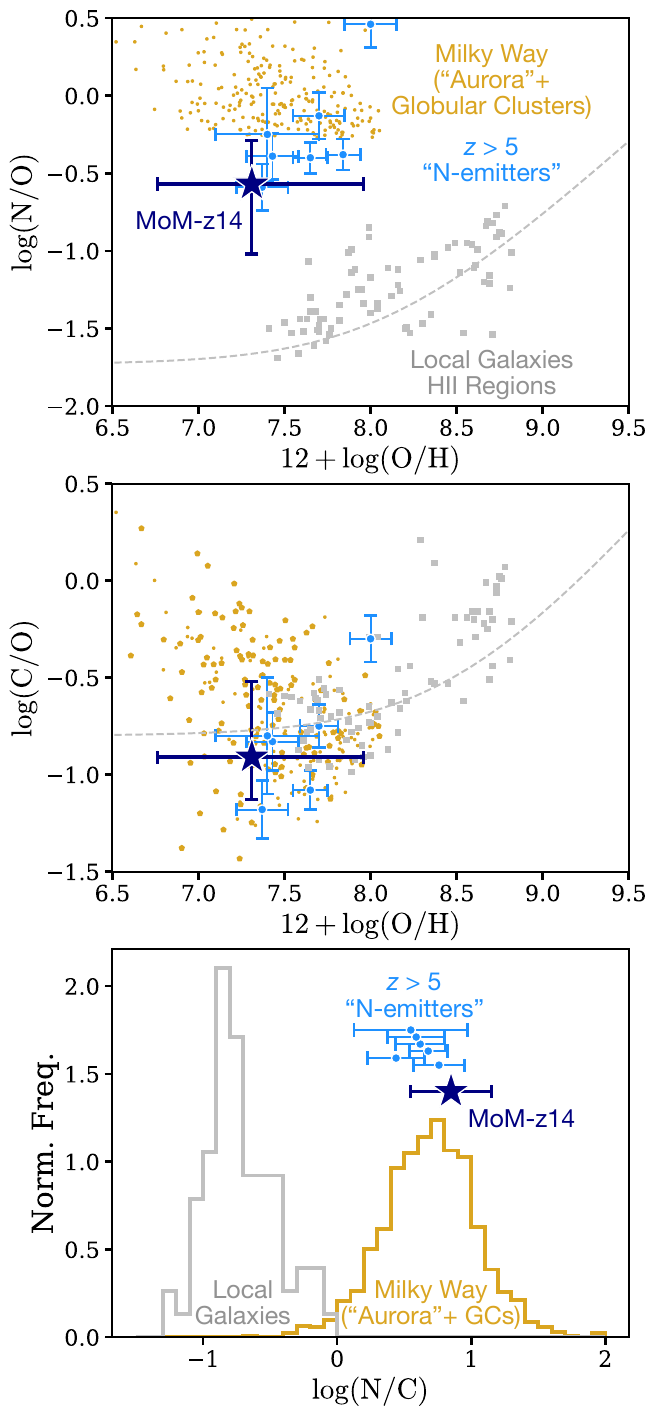}
    \caption{\textbf{The GC-like chemical abundance pattern of MoM-z14.} \textbf{Top:} MoM-z14 (navy) joins the ranks of nitrogen emitters at $z\gtrsim5$ (light blue; compilation from \citealt[][]{Schaerer24}) showing strong \nion{N}{iv}] emission and a super-solar [N/O]. For comparison we show the \citet{Nicholls17} trend that describes the locus of local star-forming galaxies (silver; \citealt[][]{Izotov23}). MoM-z14 and the N-emitters deviate by $\approx1$ dex from the local relation. Intriguingly, some of the most ancient stars born in the Milky Way at $z\gtrsim4-6$ (``Aurora"; \citealt[][]{Belokurov22, Conroy22,Rix22}) as well as local globular clusters display abundances comparable to these high-$z$ N-enriched sources (golden points; sourced from \citealt[][]{Belokurov23, Belokurov24} -- note that these stars were selected to have high $\log\rm{N/O}$). \textbf{Center:} Same as above, but for [C/O]. Here we see MoM-z14, the N-emitters, and Milky Way stars are consistent with the local scaling relation. \textbf{Bottom:} In other words, the [N/C] in these sources is highly super-solar. We note that while constraints on Oxygen are available only through a blended line, the nitrogen and Carbon lines are the brightest we measure, resulting in a relatively tight constraint (see \S\ref{sec:cue} for details).} 
    \label{fig:abundance}
\end{figure}

We deploy \texttt{Cue} \citep[][]{Li24cue} to extract insights about the nebular conditions (gas density, ionization parameter), chemical abundances ([O/H], [N/O], [C/O]), and ionizing sources ($\xi_{\rm{ion}}$) in MoM-z14. \texttt{Cue} is agnostic to the detailed physics of the source of ionizing photons powering the emission lines and parametrizes the ionizing spectrum as a flexible piecewise power-law. This is particularly useful for modeling luminous $z>10$ objects where the ionizing sources are presently unknown and may not be captured in any ab initio model. Emission lines are predicted via neural net emulation trained on \texttt{CLOUDY} \citep[][]{Ferland17}. \refrep{Note that \texttt{Cue} is designed to model emission lines, and not to perform full spectrum fitting.} Posteriors are sampled using \texttt{dynesty} \citep[][]{Speagle19}. This approach captures the complex covariance between various parameters as opposed to flattening this high-dimensional space into two-dimensional diagnostics, i.e., using specific sets of lines to bracket the density/temperature/metallicity vs. using all the lines at once to constrain all parameters simultaneously \citep[e.g.,][]{Li24GS9422}.

The results from \texttt{Cue} are summarized in Fig. \ref{fig:cue} and listed in Table \ref{tab:properties}. \refrep{In Fig. \ref{fig:cue} we have summarized the ionizing spectrum in the form of the familiar $\xi_{\rm{ion}}$ for simplicity -- in Fig. \ref{fig:cue_appendix} we show all the individual power law parameters}. Note that \nion{N}{iv} is not emulated in \texttt{Cue} and we do not use it in the fitting. All lines are satisfactorily matched within uncertainties, producing an overall $\chi^{2}/N=0.6$. We caution that the posteriors are wide and permit a broad range of outcomes. This is unsurprising given that we are working with a low-resolution spectrum where closely spaced lines are blended (e.g., the \nion{N}{iii}] quintuplet, \nion{He}{ii}+\nion{O}{iii}]), and where individual line fluxes are uncertain ($3\sigma$). \refrep{Further, posteriors for almost all quantities are truncated at the edges of the grid that \texttt{Cue} is trained on. This is an extensive grid with wider coverage than typical photoionization models \citep[e.g.,][]{Feltre16,Byler17}, and yet extreme sources like MoM-z14 may require even broader sets of parameters.} Nonetheless, the results in Fig. \ref{fig:cue} give us a first order, preliminary portrait of the ISM in MoM-z14.

\refrep{The \texttt{Cue} fits that take into account the observed EWs and line ratios suggest an efficiently ionizing source -- $\log(\xi_{\rm{ion}}/\rm{erg^{-1}\ \rm{s^{-1}}})=26.3\pm0.5$ --  irradiating the gas with an ionization parameter of $\log(U)\approx-1.5$. Within errors, this is consistent with star-forming galaxies observed at $z>6$ \citep[e.g.,][]{Simmonds24}, with Very Massive/Super Massive Stars that lie around the stellar population maximum $\log(\xi_{\rm{ion}}/\rm{erg^{-1}\ \rm{s^{-1}}})\approx26$ \citep[e.g.,][]{Schaerer25VMS}, as well as metal-poor AGN \citep[e.g.,][]{Nakajima22}. The relatively uncertain measurements currently in hand are unable to directly isolate an ionizing source across these scenarios – the \ion{He}{2}, \ion{C}{3}], \ion{C}{4}, \ion{N}{4}], \ion{N}{3}] EWs are all consistent with metal-poor AGN as well as star-forming galaxies. Indeed, \citet[][]{Calabro24} have shown in detail the difficulty of distinguishing between these populations at PRISM resolution using EWs of these emission lines in their analysis of GLASS-z12/GHz2. However, as we discuss later here and in \S\ref{sec:gcs}, the chemical abundances, especially [N/C] are relatively robust among all our inferred quantities and may already yield insights on the ionizing sources.}

The inferred gas density is $\approx20\times$ higher than JADES-GS-z14-0, which was found to be gas depleted perhaps due to intense feedback \citep[e.g.,][]{Schouws25CII,Carniani25CII}. Resolving the density sensitive \nion{N}{iii}], \nion{N}{iv}], and \nion{C}{iii}] lines with $R\approx1000$ spectra will help us precisely constrain exactly how high the gas density is. \refrep{For now we note that while high ionization parameters ($\log(U)>-1$) and electron densities ($n_{\rm{H}}$> $10^3$ cm$^{-3}$) might seem extreme, such values are plausible for high-redshift galaxies \citep[e.g.,][]{Topping24} and may even indicate AGN activity \citep[e.g.,][]{Maiolino24GNz11}.}

The derived chemical abundance patterns are contextualized in Fig. \ref{fig:abundance}. We find evidence for super-solar [N/O] at relatively low [O/H] of $\approx-1.3$, while the [C/O] is as expected based on local scaling relations. Though note that the [N/O] from \texttt{Cue} may be underestimated given that we do not include \nion{N}{iv}]. To derive an independent estimate of [N/C], we follow \citet{Villar-Martin04} who performed a detailed analysis of the nitrogen-emitting Lynx arc \citep[][]{Fosbury03}, a lensed galaxy at $z=3.357$. In particular, we can bypass the highly uncertain Oxygen abundance (which we have constraints on only through a blend) and directly derive N/C $\approx$ (N$^{3+}$+N$^{2+}$)/(C$^{3+}$+C$^{2+}$) where these ionic abundances follow from \nion{N}{IV}], \nion{N}{III}], \nion{C}{IV}, and \nion{C}{III}]. The free parameter is the electron temperature, for which we explore a wide range ($T_{\rm{e}}/(10^{4}\rm{K})=0.5-3$) bracketing typical values inferred for comparable $z>10$ sources \citep[e.g.,][]{Cameron23gnz11,Calabro24,Carniani25CII}. This translates to values of $\log(\rm{N/C})=0.7^{+0.15}_{-0.15}$ to $1.1^{0.15}_{-0.15}$, which maps to $\rm{[N/C]}=1.3^{+0.15}_{-0.15}$ to $1.7^{+0.17}_{-0.17}$. That is, pending confirmation with higher resolution, higher SNR spectroscopy, MoM-z14 may be among the most nitrogen enhanced galaxies discovered with JWST yet with a $>10\times$ enhancement relative to the solar abundance.

Such nitrogen enrichment has been reported in a wide range of objects with JWST -- two of the brightest sources at $z>10$ (GNz11, GLASS-z12/GHz2; e.g., \citealt[][]{Cameron23gnz11,Castellano24}), some Little Red Dots and broad-line AGN \citep[e.g.,][]{Ubler23, Labbe24, Isobe25}, and half a dozen individual star-forming galaxies \citep[e.g.,][]{Isobe23, Topping24, Marques-Chaves24, Schaerer24}. Prior to JWST, only a handful of such strong N-emitting galaxies had been reported across the entire extragalactic observational literature \citep[][]{Fosbury03, Patricio16,Mingozzi22, Pascale23}. Like MoM-z14, most of these sources are extremely compact relative to their peers at similar redshift, perhaps suggesting shared origins \citep[e.g.,][]{Harikane24, Schaerer24}. 

The key to unraveling this shared origin may lie in Galactic archaeology -- elevated [N/C] comparable to these sources is routinely observed in Milky Way globular clusters (GCs) as well as some of the most ancient, metal-poor stars born in the Galaxy \citep[e.g.,][]{Belokurov23, Belokurov24}. MoM-z14 and the range of objects discussed above may be the ``live action" versions of these dense massive clusters and early epochs of star-formation -- we discuss this line of reasoning further in \S\ref{sec:gcs}.

\subsection{An Absent Damping Wing at $z_{\rm{spec}}=14.44$?} \label{sec:Lya}

\begin{figure}
    \centering
    \includegraphics[width=\linewidth]{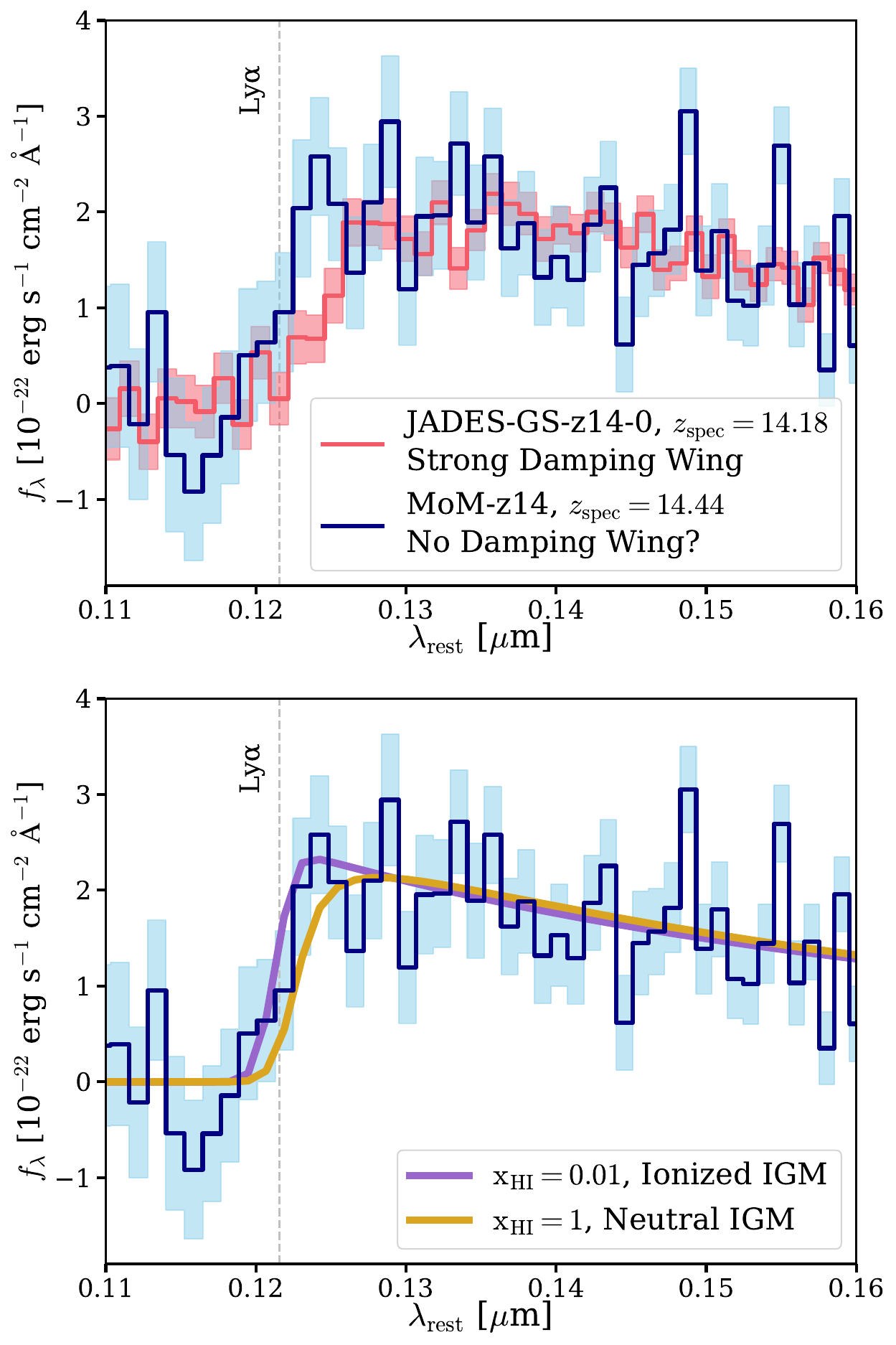}
    \caption{\textbf{A dearth of neutral gas and an ionized IGM around MoM-z14?} \textbf{Top:} The relatively sharp Ly$\alpha$ break of MoM-z14 is contrasted against the UV ``turnover" seen in JADES-GS-z14-0 (red; \citealt{Carniani24}). This turnover is routinely observed in $z\gtrsim8$ galaxies \citep[e.g.,][]{Heintz24} and has been interpreted as damped Ly$\alpha$ absorption due to dense columns of neutral gas -- $\log( N_{\rm{HI}}/\rm{cm}^{-2})\gtrsim10^{22}$ -- ensconcing these galaxies \citep[e.g.,][]{ Heintz25_GSz14}. MoM-z14 and its immediate environment may lack such neutral gas reservoirs. For the comparison above, the JADES-GS-z14-0 spectrum has been normalized to the $M_{\rm{UV}}$ of MoM-z14. The spectra are reduced similarly with the \texttt{msaexp} software following the reduction choices of the DAWN JWST Archive (DJA v4). Note that both galaxies have precise systemic redshifts measured from emission lines enabling this comparison. \textbf{Bottom:} Best-fit models for the Ly$\alpha$ region in MoM-z14 assuming a highly ionized ($\rm{x}_{\rm{HI}}=0.01$, purple) and fully neutral ($\rm{x}_{\rm{HI}}=1$, gold) IGM with no DLA included and the remaining parameters varied 
    (i.e., the shape and normalization of the UV continuum -- $\beta_{\rm{UV}}$ and $F_{\rm{0}}$). The sharp break in MoM-z14 may disfavor an entirely neutral IGM in its immediate vicinity.}
    \label{fig:dla}
\end{figure}

\begin{figure}
    \centering
\includegraphics[width=\linewidth]{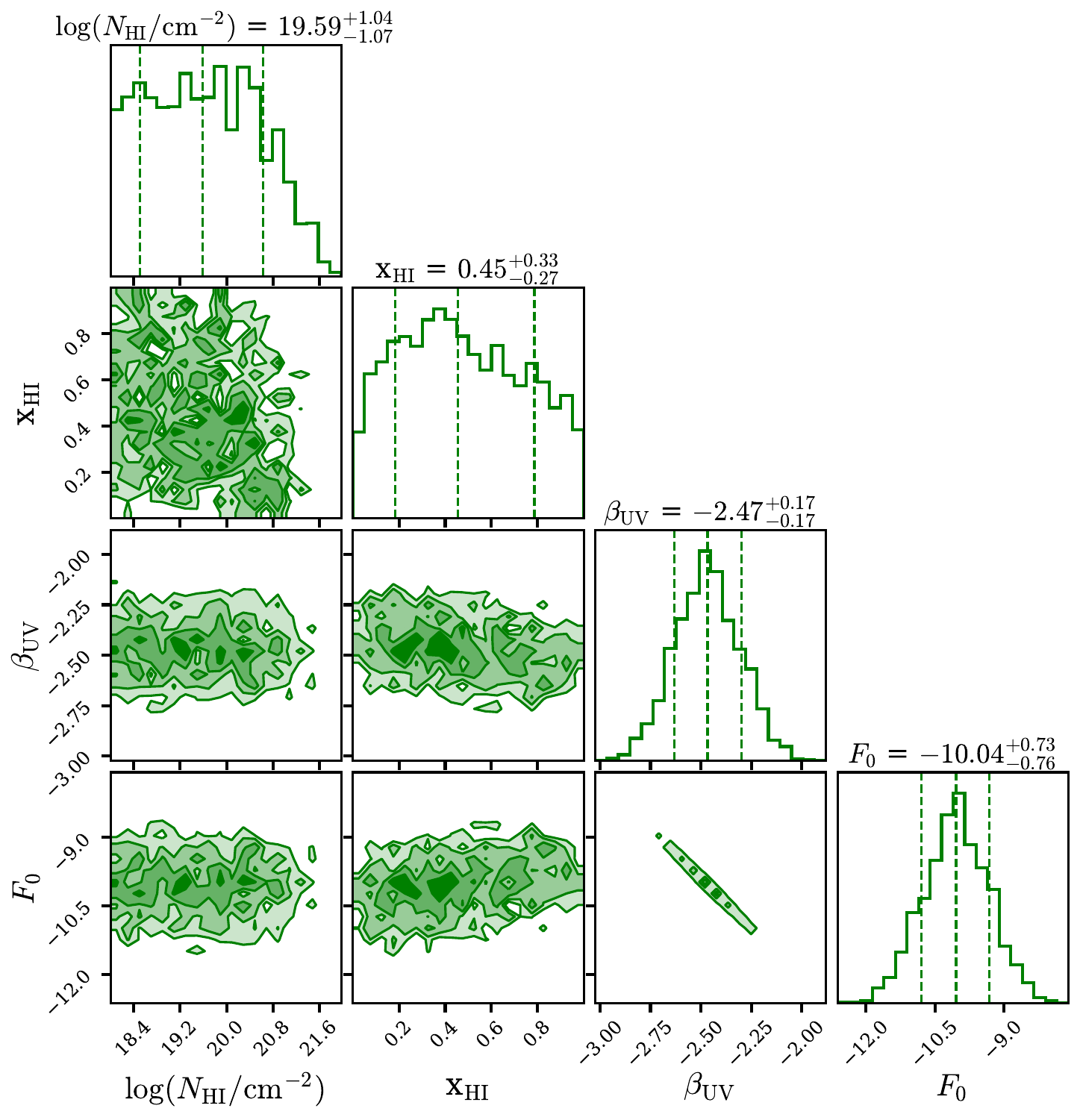}
    \caption{\refrep{\textbf{Constraints on the DLA and IGM neutral fraction in the vicinity of MoM-z14}. The redshift is held fixed in this model at $z=14.44$ based on the detection of multiple UV lines (\S\ref{sec:specz}). The inferred column density of neutral gas, $\log(N_{\rm{HI}}/\rm{cm}^{-2})$, in MoM-z14 is more than an order of magnitude lower than in JADES-GS-z14-0 ($22.27^{+0.08}_{-0.09}$; \citealt[][]{Heintz25_GSz14}) supporting the comparison in Fig. \ref{fig:dla} suggesting the lack of a DLA. Furthermore, the inferred IGM neutral fraction ($x_{\rm{HI}}$) is not $\approx100\%$ -- while the posteriors are broad, $x_{\rm{HI}}<90\%$ is favored at $>93\%$ significance with the most likely value being $\approx40\%$. Deeper, higher resolution data (e.g., to rule out Ly$\alpha$) is required to confirm this hint of an ionized region at such a high redshift. $\beta_{\rm{UV}}$ and $F_{\rm{0}}$ are parameters describing the shape and normalization of the spectrum (\S\ref{sec:lyaz}).} }
    \label{fig:xhi}
\end{figure}

\begin{figure*}
    \centering
\includegraphics[width=\linewidth]{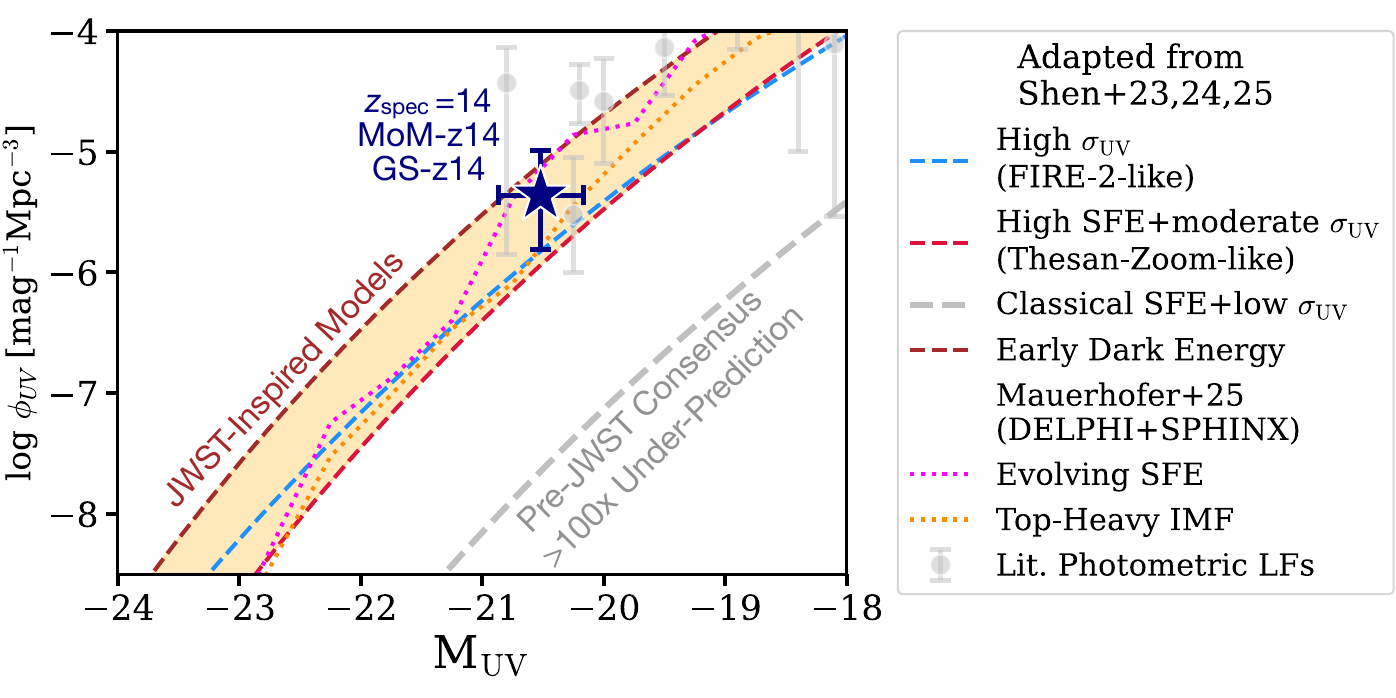}
    \caption{\textbf{The $z_{\rm{spec}}=14-15$ UV luminosity function (LF; navy blue star) based on MoM-z14 and JADES-GS-z14-0 favors the latest JWST-inspired theoretical models (brown shaded area) and a dramatic departure from pre-JWST consensus LFs (silver dashed)}. While at $z\approx10-12$ the tension with the pre-JWST models is on the $\approx10\times$ level, at $z\approx14-15$ it is much more pronounced ($>100\times$). Theoretical LFs typifying various physical scenarios that may explain the abundance of bright galaxies are illustrated based on \citet[][]{Shen23,Shen24,Shen25} and \citet[][]{Mauerhofer25} whereas photometric LFs are shown for comparison from \citet{Finkelstein24, Robertson24, Donnan24primer, Whitler25, Morishita24}. These scenarios, each with distinct merits, ultimately invoke some combination of: (i) a higher star-formation efficiency that marks the rate at which gas cools into stars (SFE); (ii) a higher UV variability ($\sigma_{\rm{UV}}$) that captures phenomena such as bursty star-formation; (iii) the presence of luminous ingredients that preferentially occur in the early Universe (e.g., a top-heavy IMF, AGN); (iv) modifications to cosmology that increase the abundance of massive dark matter halos in the early universe (e.g., early dark energy). What is also clear from this figure is that merely measuring the UVLF is insufficient to distinguish between these scenarios to reveal what makes these galaxies shine so bright -- detailed studies of individual sources, as well as supporting population level statistics (e.g., clustering) are required.}
    \label{fig:uvlf}
\end{figure*}

A puzzling discovery from early observations with JWST was a set of sources with apparent strong UV ``turnovers'' where the Ly$\alpha$ break appeared offset from the systemic redshift. This was first interpreted as extremely strong damped Ly$\alpha$ absorption (DLA) due to dense H\,{\sc i} gas in the vicinity of these galaxies by \citet{Heintz24_DLA}. This phenomenon is now frequently observed among the most distant galaxies at $z=10-14$ \citep{Umeda24,Hainline24,DEugenio24,Asada24,Witstok25,Heintz24,Heintz25_GSz14}. Contributions from strong nebular continuum and 2-photon emission have also been argued to explain at least a subset of galaxies with strong UV turnovers \citep{Cameron24,Katz24}. However, this interpretation has been excluded for the vast majority of high-$z$ galaxies based on the lack of strong emission-line EWs or too blue rest-UV slopes \citep[][Pollock et al. in prep.]{Chen24,Witstok25}. 

In the top panel of Fig. \ref{fig:dla} we show qualitatively that MoM-z14 has a relatively sharp Ly$\alpha$ break and does not show a broad UV turnover. We contrast the source with JADES-GS-z14-0 at a slightly lower redshift that displays a DLA profile corresponding to $\log(N_{\rm{HI}}/\rm{cm^{-2}})\approx 22.3$ \citep[][]{Heintz25_GSz14,Carniani24}. Not only is such a DLA apparently absent in MoM-z14, but even the IGM around it may be partially ionized. The bottom panel of Figure~\ref{fig:dla} shows Ly$\alpha$ damping wing curves for the IGM with neutral fractions fixed to $\rm{x}_{\rm HI} = 0.01$ and 1.0 (i.e. 1\% to 100\% neutral). Intriguingly, this comparison hints that MoM-z14 may reside in a partially ionized IGM (see also the recent discovery of Ly$\alpha$ at $z>10$ in \citealt[][]{Bunker23} and \citealt[][]{Witstok25}). We also note that in contrast to JADES-GS-z14-0, MoM-z14's blue UV slope and inferred hard ionizing spectrum are telltale signs of a non-zero Lyman continuum escape fraction, consistent with it contributing to the ionization of its surroundings, helping explain the difference in the damping signatures between these two sources \citep[e.g.,][]{Chisholm22, NM22, Jaskot24pt1, Jaskot24pt2}. 

\refrep{To quantify these insights, we fit a model similar to the one discussed in \S\ref{sec:lyaz} but with the redshift fixed to the systemic UV line redshift ($z=14.44$) and an extra parameter ($N_{\rm{HI}}$) so we may account for the local, dense neutral gas (i.e., the DLA) in addition to the neutral fraction of the IGM ($x_{\rm{HI}}$). The DLA adds an additional Ly$\alpha$ absorption component to the power-law+IGM model introduced in Sect.~\ref{sec:lyaz} in the form of a Voigt profile, using the approximation from \citet{TepperGarcia06}.} 

\refrep{In Fig. \ref{fig:xhi} we present the results of the fit to this model. The posteriors are broad, but bear intriguing hints of an early ionized region. A strong DLA as seen in JADES-GS-z14-0 ($\log(N_{\rm{HI}}/\rm{cm}^{-2}=22.27^{+0.08}_{-0.09}$) and typical $z>10$ galaxies ($\log(N_{\rm{HI}}/\rm{cm}^{-2}\gtrsim21$; \citealt[][]{Heintz24}) is disfavored at $>92\%$ confidence. The IGM neutral fraction $x_{\rm{HI}}$ is inferred to be $0.45^{+0.33}_{-0.27}$ with a completely neutral sight-line disfavored at $>93\%$ confidence. We emphasize caution that at the moment we are unable to constrain the strength of Ly$\alpha$ or \nion{N}{V} emission, which at this low resolution and SNR makes detailed interpretation of the break challenging. However, we hope these preliminary results motivate the community to undertake the intensive spectroscopy required to definitively collapse these posteriors.}

\refrep{It is interesting to note that even a modest $\approx10-20\%$ ionized fraction at $z>10$ may help alleviate some of the most pressing tensions in $\Lambda$CDM cosmology (e.g., the problem of negative neutrino masses and hints of evolving dark energy) by providing support for a higher value of the Thomson optical depth ($\tau$) compared to the value inferred by \citet[][]{Planck18} \citep[e.g.,][]{Giare24, Asthana24, Mason25, Sailer25}.}



\section{Discussion}

\subsection{The spectroscopic UV luminosity function at $z_{\rm{spec}}\approx14-15$}
\label{sec:uvlf1}
The bright end of the UV luminosity function (UV LF) at $z>10$ has been a key benchmark for early Universe models in the JWST era. The chasm between theory and observations suggested by GN-z11 \citep{Oesch16} at $z\approx10$ was immediately borne out by photometric samples \citep[e.g.,][]{Naidu22,Castellano22, Donnan23,Harikane23,Finkelstein23, Adams24} and subsequently confirmed with spectroscopy \citep[e.g.,][]{ArrabalHaro23, Castellano24, Napolitano25}. The discovery of JADES-GS-z14-0 in a $\approx10$ arcmin$^{2}$ field hinted that bright sources remained abundant even at higher redshifts, where the tension with models may be even more pronounced \citep[e.g.,][]{Robertson24, Whitler25,PerezGonzalez25_superhighz, Castellano25_superhighz}. 

Here we confirm this picture. The number density implied by MoM-z14 and JADES-GS-z14-0 is shown in Fig. \ref{fig:uvlf}. To estimate this LF point we assume MoM-z14 and JADES-GS-z14-0 are the only $-21<M_{\rm{UV}}<-20$ sources over the UDS, COSMOS, and GOODS-S fields at $z=14-15$ in the area imaged by the PRIMER and JOF surveys \citep[][]{Donnan24primer,Eisenstein23jof}. \refrep{The survey area used for this calculation is 136 (PRIMER-COSMOS; \citealt[][]{Donnan24primer}) + 243 (PRIMER-UDS; \citealt[][]{Donnan24primer}) + 9 arcmin$^{2}$ (JOF; \citealt[][]{Robertson24}). The differential co-moving volume at $z=14-15$ for this survey area is 462,648 Mpc$^{3}$.}. Our derived number density -- $\log(\phi/\rm{Mpc^{-3}\ \rm{mag}^{-1} }) = -5.36^{+0.37}_{-0.45}$ -- is in excellent agreement with photometric luminosity functions recently reported at $z\approx14-15$ \citep[e.g.,][]{Finkelstein24, Donnan24primer,Robertson24,Whitler25}.

To represent the pre-JWST theoretical consensus, we construct a $\Lambda$CDM UV LF that assumes a literature-averaged star-formation efficiency (SFE), low UV variability ($\sigma_{\rm{UV}}$), and a \citet{Chabrier03} IMF following the framework outlined in \citet{Shen23}. Such an LF underestimates the incidence of observed sources by $\approx200\times$ ($182^{+329}_{-105}\times$) and is disfavored at high confidence. A similar level of over-abundance is seen relative to individual pre-JWST literature LFs \citep[e.g.,][]{Mason15, Tacchella18, Behroozi19, Dayal22}. This comparison makes clear that the new generation of models being developed in response to early JWST observations is very much required.

\subsection{What makes luminous $z>10$ galaxies shine so bright?}
\label{sec:uvlf2}

In Fig. \ref{fig:uvlf} we show luminosity functions representing various classes of solutions proposed to address the abundance of bright galaxies at $z>10$. To provide a controlled comparison, we illustrate these solutions with a handful of models, keeping all else fixed \citep{Shen23,Shen24,Shen25, Mauerhofer25}. Broadly speaking, these solutions invoke some combination of the following: 

\textit{(i) A higher UV variability} ($\sigma_{\rm{UV}}$), which represents the scatter in UV brightness at fixed mass. The key form of $\sigma_{\rm{UV}}$ explored in the literature has been ``bursty" star-formation, where galaxies oscillate between periods of quiescence and starbursts on short $\approx10$ Myr timescales \citep[e.g.,][]{Faucher-Giguere18,Mason23, Sun23bursty, Nikopoulos24, Endsley24bursty, Looser24, Kravtsov24}. While indeed, galaxies like MoM-z14 show a rising SFH, constraining the magnitude of this rise purely from rest-UV SEDs is challenging (see the posteriors in Fig. \ref{fig:prosp}; c.f. \citealt[][]{Harikane2025UVLF, Kokorev25}). We also note that the $\sigma_{\rm{UV}}$ required to match the $z_{\rm{spec}}\approx14-15$ LF is much higher, $\approx1.5-2$ mag \citep[e.g.,][]{Shen23, Shen24}, compared to the $\approx0.7$ mag inferred out to $z\approx9$ from galaxy clustering \citep[e.g.,][]{Shuntov25}.

\textit{(ii) A higher star-formation efficiency (SFE)}, representing the conversion rate between gas and stars. For example, \citet{Dekel23, Dekel25, Li24} envision feedback-free star-formation wherein dense gas ($n_{\rm{H}}\gtrsim10^{3} \rm{cm^{-3}}$) at low metallicity ($\lesssim20\% Z_{\rm{\odot}}$) may collapse into stars at a rate much faster than can be regulated by supernovae or stellar winds (see also \citealt[][]{Boylan-Kolchin25}). These ISM conditions are not implausible in light of ``black hole stars" with $n_{\rm{H}}\approx10^{10} \rm{cm^{-3}}$ \citep[e.g.,][]{Naidu25BHstar,degraaff25} thought to power Little Red Dots -- LRDs occur at a similar rate as the luminous $z>10$ galaxies ($\approx10^{-5}\ \rm{cMpc^{-3}}$; e.g., \citealt[][]{Matthee24, Akins24, Kokorev24}). Indeed, MoM-z14 does show hints of the required ISM conditions for this scenario (Fig. \ref{fig:cue}) -- this can be further tested with high-resolution spectroscopy to resolve density-sensitive multiplets (e.g., \nion{N}{iii}]; \citealt{Maiolino24GNz11}).

\textit{(iii) Modifications to $\Lambda$CDM} that enhance the abundance and accretion rate of early halos \citep[e.g.,][]{Liu22,Huang23, Parashari23, Gouttenoire24}. For instance, in Early Dark Energy (EDE), an increased early expansion relative to $\Lambda$CDM decreases the physical sound horizon measured in CMB, resulting in shifts of the inferred cosmological parameters. This solves the Hubble tension with a side effect of enhancing dark matter halo abundance at high redshifts (see \citealt[][]{Poulin23} for a recent review). Clustering is a promising test of whether luminous galaxies are resident in truly massive halos \citep[e.g.,][]{Munoz23,Gelli24,Eilers24} -- in the EDE scenario one would expect a higher halo mass on average relative to say $\Lambda$CDM with a high $\sigma_{\rm{UV}}$. At the moment, we do not find any obviously associated $z_{\rm{phot}}=14.44$ sources around MoM-z14, but much deeper imaging and spectroscopy will be required to rule out any companions. No spectroscopically associated sources or overdensities have been reported for other luminous $z>10$ sources either (though see \citealt[][]{Tacchella23GNz11} for possible photometric companions around GNz11).

\textit{(iv) Boosting the light to mass ratio relative to lower-redshift galaxies}. Examples include decreasing dust attenuation with redshift \citep[e.g.,][]{Ferrara23}, a top-heavy IMF \citep[e.g.,][]{Yung24, Hutter25,Lu25}, or a high incidence of AGN \citep[e.g.,][]{Maiolino24GNz11,Calabro24, Natarajan24}. MoM-z14 does show hints of a blue $\beta_{\rm{UV}}$ slope consistent with little dust, and a chemical abundance pattern consistent with very massive stars that may emerge under a top-heavy IMF (see \S\ref{sec:gcs}). The presence of strong UV lines in the prism spectrum guarantee that higher resolution spectroscopy will be fruitful in drastically shrinking the posteriors in Fig. \ref{fig:cue} (e.g., of $\xi_{\rm{ion}}$ and [N/C]) to test these scenarios.

\subsection{Insights from the Milky Way into MoM-z14, early bright galaxies, and supermassive black holes}
\label{sec:gcs}


Insight into the underlying physics of strong nitrogen emitters may be found in the Milky Way, where nitrogen enhancement is a characteristic feature of globular cluster (GC) stars -- see \citet[][]{Bastian18} and \citet{Gratton19} for recent reviews. 
Furthermore, super-solar nitrogen is also observed in 2--4\% of the most metal-poor stars that formed in the Milky Way prior to the formation of the disk(s) at $z\gtrsim4-6$ (the ``Aurora" or ``proto-galaxy" component of the Milky Way; \citealt{Conroy22, Rix22, Belokurov22, Chandra24, Semenov24howearly}). \refrep{Note that Aurora is a distinct in-situ (born in the Milky Way) component of the stellar halo, which is different from the majority of halo stars that arise from the debris of accreted dwarf galaxies \citep[e.g.,][]{Belokurov18,Helmi18, Naidu20halo}}. 
By using nitrogen as a proxy for formation in dense clusters, $\approx50-70\%$ of star-formation at $z\gtrsim4-6$ in the Milky Way (i.e., in Aurora) is estimated to have occurred within such bound structures \citep[e.g.,][]{Horta21, Belokurov23}. Intriguingly, the fraction of N-enhanced stars then rapidly falls off towards higher metallicities ($z\lesssim4$) by few orders of magnitude as the Milky Way's disk begins spinning up \citep[e.g.,][]{Belokurov23, Belokurov24, Kane25}. Indeed, direct observations of lensed galaxies at $z\gtrsim6$ are direct reflections of this archeological inference \citep[e.g.,][]{Fujimoto24GCs, Adamo24}. Therefore, the N-enhancement in MoM-z14 and other high-$z$ sources seems far from peculiar, and perhaps reflects the generic mode of star-formation in very dense clusters prevalent at these epochs \citep[e.g.,][]{Ma20GCs}.

As for the underlying physical picture, these primordial dense clusters may be hotbeds for peculiar stellar populations and unique ISM conditions. For example, the rate of rapid runaway collisions in massive GCs is expected to be significant, and may help produce ``very massive stars" (VMS, $\approx100-1000 M_{\rm{\odot}}$; e.g., \citealt[][]{Vink24}) or ``supermassive stars" (SMS, $\approx10^4 M_{\rm{\odot}}$; e.g., \citealt[][]{Denissenkov14, Gieles18, Vergara25}). These stars would then leave behind an enriched ISM with characteristic chemical abundance patterns (from e.g., hot H burning in the CNO cycle; \citealt[][]{Charbonnel23}). 

Further, beyond abundances, with objects like SMS and VMS, the top-heavy IMF and drastically altered mass-to-light ratio \citep[e.g.,][]{Martins20} would help reconcile theoretical and observed UV LFs. The $M_{\rm{\star}}\approx10^{8} M_{\rm{\odot}}$ we report for MoM-z14 then may be an upper limit on the stellar mass, implying $\lesssim10$ massive ($\approx10^{6-7} M_{\rm{\odot}}$) GCs may account for all the light. This may also be linked to the compactness of N-emitters -- if only a few GCs dominate the light, they may produce a light profile with a very small effective radius (Fig. \ref{fig:bimodal}). The stellar surface density we infer in MoM-z14 ($\Sigma_{*}\approx10^{9-10}$ $M_{\rm{\odot}}\ \rm{kpc^{-2}}$) is comparable to Milky Way GCs, even if the stellar mass is over-estimated by 1-2 dex \citep[e.g.,][]{Hopkins10}. The hard ionizing spectrum (\S\ref{sec:cue}) could also be naturally explained by an IMF that is dominated by these highly ionizing massive stars \citep[e.g.][]{martins25}. What then of the extended sources such as JADES-GS-z14-0 comprising the upper branch in Fig. \ref{fig:bimodal}? These sources may represent a mode of star-formation corresponding to the $\approx50\%$ of nitrogen-poor stars forming in the Milky Way in these early epochs.

As for the connection between UV-bright galaxies, N-rich LRDs \citep[e.g.,][]{Labbe24}, and broad-line AGN \citep[e.g.,][]{Isobe25}, perhaps the remains of VMS and SMS mark the beginnings of gas-enshrouded supermassive black holes (``black hole stars") that form deeply embedded in this dense environment \citep[e.g.,][]{Ji25BT, Naidu25BHstar, degraaff25, Rusakov25, Taylor25}. It is tempting to speculate that the drop in LRD number-density by an order of magnitude below $z\approx4$ \citep[e.g.,][]{Ma25,Bisigello25} is linked to the similar drop in the occurrence of N-enriched stars in the Milky Way at these redshifts. These promising, yet preliminary connections between the highest redshift galaxies and the local archaeological record hinge on detailed chemical abundances, motivating investments in deep, high-resolution spectroscopy.

\section{Summary \& Outlook}

This paper presented first results from the ``Mirage or Miracle" (MoM) survey that we designed to spectroscopically test the abundance and nature of luminous galaxy candidates at $z>10$. Here we introduced MoM-z14, a remarkably bright ($M_{\rm{UV}}=-20.2$) source at $z_{\rm{spec}}=14.44$ whose discovery pushes the cosmic frontier to $\approx280$ million years after the Big Bang. We find the following:

\begin{itemize}

    \item The NIRCam imaging of MoM-z14 shows a robust dropout signature ($z_{\rm{phot}}=14.86^{+0.47}_{-1.50}$), which the follow-up NIRSpec prism spectrum reveals is due to a sharp Ly$\alpha$ break at $z_{\rm{break}}=14.42^{+0.10}_{-0.09}$. Remarkably for a galaxy at this early epoch, the prism spectrum displays multiple rest-UV emission lines (\nion{C}{iv}, \nion{C}{iii}], \nion{N}{iv}, \nion{N}{iii}, \nion{He}{ii}+\nion{O}{iii}]). This allows for a rather precise redshift determination ($z_{\rm{UV\ lines}}=14.44^{+0.02}_{-0,02}$) while also providing a unique opportunity for detailed physical characterization. 
    [Figs. \ref{fig:data}, \ref{fig:specz}, Table \ref{tab:lines}, \S\ref{sec:specz}]
    
    \item The source is quite compact, and yet spatially resolved, thereby disfavoring a dominant AGN contribution (circularized $r_{\rm{e}}=74^{+15}_{-12}$ pc). It joins GNz11 and GLASS-z12/GHz2 as an outlier in being extremely compact for its luminosity and redshift. [Figs. \ref{fig:morphology}, \ref{fig:bimodal}, \S\ref{sec:morphology}]

    \item  Modeling the SED and strong UV emission lines reveals an SMC-like dwarf galaxy ($\approx10^{8} M_{\rm{\odot}}$) caught in a burst, efficiently emitting copious ionizing photons through a virtually dust-free ISM [Figs. \ref{fig:prosp}, \ref{fig:cue}, Table \ref{tab:properties}, \S\ref{sec:prosp}, \S\ref{sec:cue}]
    
    \item Perhaps relatedly, the immediate surroundings of MoM-z14 appear to be partially ionized as borne out by the absence of a strong damping wing. Ly$\alpha$ follow-up and deeper spectroscopy of the break shape will help determine if we may be witnessing an earlier than expected start to reionization. [Fig. \ref{fig:dla}, \S\ref{sec:Lya}]

    \item We report the bright-end of the spectroscopic UV LF at $z\approx14-15$ based on MoM-z14 and JADES-GS-z14-0 ($z_{\rm{spec}}=14.18$). We confirm that the $>100\times$ over-abundance relative to pre-JWST consensus models suggested by photometric samples is not a mirage. We demonstrate four classes of model solutions to bridge this chasm (UV variability, higher SFE, cosmology, modifying the mass to light ratio). The UV LF alone cannot discriminate among these solutions and deep follow-up observations of individual sources and their environments are required. [Fig. \ref{fig:uvlf}, \S\ref{sec:uvlf1},\S\ref{sec:uvlf2}]
    
    \item MoM-z14 is a strong \nion{N}{iv}]$\lambda1487$\AA\ emitter, adding the highest redshift example yet to this emerging class of sources that now includes a collection of luminous Little Red Dots, broad-line AGN, and extremely compact star-forming galaxies. In fact, pending confirmation with high-resolution spectroscopy, MoM-z14 may rank among the most nitrogen-enhanced sources discovered with JWST yet ([N/C]$>1$). It adds further evidence for a size-chemistry bimodality at $z>10$, wherein extended sources tend to be nitrogen weak while compact sources are strong N emitters. [Figs. \ref{fig:bimodal}, \ref{fig:abundance}, \S\ref{sec:cue}]
    
    \item We interpret MoM-z14 and N-emitters through Galactic archaeology, connecting their abundance patterns to the most ancient stars born in the Milky Way at $z\gtrsim4$ (``Aurora"/the ``proto-galaxy") as well as to globular clusters. The N-enhancement, brightness, hard ionizing spectra, stellar density, morphology, redshift dependence, and black hole fraction of these sources may be linked to GC-like environments wherein runaway collisions may produce extraordinary objects such as supermassive stars. [Figs. \ref{fig:bimodal}, \ref{fig:abundance}, \S\ref{sec:gcs}]
    
\end{itemize}

That the number density of luminous galaxies evolves only gradually between $z\approx10$ and $z\approx14-15$ is now on firm spectroscopic footing. The good fortune of inhabiting a Universe teeming with GN-z11s means remarkably luminous $z\approx15$ galaxies in the hundreds  may be within the grasp of the Roman Space Telescope. JWST itself appears poised to drive a series of great expansions of the cosmic frontier -- previously unimaginable redshifts, approaching the era of the very first stars, no longer seem far away.

\section*{Acknowledgments}

\refrep{We thank the two anonymous referees for their insightful comments that have strengthened this work.} ``Mirage or Miracle" is but the latest link in a long chain of surveys that have built COSMOS into a premier extragalactic legacy field. We are thankful to all the teams who have contributed to this legacy, particularly those mentioned in \S\ref{sec:data} for leading recent JWST programs whose imaging we have incorporated in our analysis. We are grateful to Vasily Belokurov for help in compiling the Milky Way reference sample featured in Fig \ref{fig:abundance}. We thank Danielle Berg for sharing a highly complete, highly decimalized NUV vacuum line list. We are grateful to our program's NIRSpec reviewer, Dan Coe, and program coordinator, Allison Vick, for valuable input on our MSA design. We acknowledge illuminating conversations with Risa Wechsler and Chao-Lin Kuo about early reionization. RPN thanks Neil Pappalardo and Jane Pappalardo for their generous support of the MIT Pappalardo Fellowships in Physics, and for their enthusiasm and encouragement for seeking galaxies at the highest redshifts. 

RPN acknowledges funding from {\it JWST} program GO-5224. Support for this work was provided by NASA through the NASA Hubble Fellowship grant HST-HF2-51515.001-A awarded by the Space Telescope Science Institute, which is operated by the Association of Universities for Research in Astronomy, Incorporated, under NASA contract NAS5-26555. 
This work has received funding from the Swiss State Secretariat for Education, Research and Innovation (SERI) under contract number MB22.00072, as well as from the Swiss National Science Foundation (SNSF) through project grant 200020\_207349.
Funded by the European Union (ERC, AGENTS, 101076224 and HEAVYMETAL, 101071865). Views and opinions expressed are however those of the author(s) only and do not necessarily reflect those of the European Union or the European Research Council. Neither the European Union nor the granting authority can be held responsible for them.  
The Cosmic Dawn Center (DAWN) is funded by the Danish National Research Foundation under grant DNRF140.
This work has also been supported by JSPS KAKENHI Grant Number 23H00131. HA acknowledges support from CNES, focused on the JWST mission, and the Programme National Cosmology and Galaxies (PNCG) of CNRS/INSU with INP and IN2P3, co-funded by CEA and CNES. HA is supported by the French National Research Agency (ANR) under the project FIRSTGAL, grant number ANR-24-CE31-0838. SB is supported by the UK Research and Innovation (UKRI) Future Leaders Fellowship [grant number MR/V023381/1]. R.D.~acknowledges support from the INAF GO 2022 grant ``The birth of the giants: JWST sheds light on the build-up of quasars at cosmic dawn'' and by the PRIN MUR ``2022935STW'', RFF M4.C2.1.1, CUP J53D23001570006 and C53D23000950006.

Computations supporting this paper were run on MIT's Engaging cluster. This publication made use of the NASA Astrophysical Data System for bibliographic information. Some of the data products presented herein were retrieved from the Dawn JWST Archive (DJA). DJA is an initiative of the Cosmic Dawn Center (DAWN), which is funded by the Danish National Research Foundation under grant DNRF140. Software used in developing this work includes: \texttt{matplotlib} \citep{matplotlib}, \texttt{jupyter} \citep{jupyter}, \texttt{IPython} \citep{ipython}, \texttt{numpy} \citep{numpy}, \texttt{scipy} \citep{scipy}, \texttt{TOPCAT} \citep{topcat}, and \texttt{Astropy} \citep{astropy}.

This work is based on observations made with the NASA/ESA/CSA James Webb Space Telescope. The data were obtained from the Mikulski Archive for Space Telescopes at the Space Telescope Science Institute, which is operated by the Association of Universities for Research in Astronomy, Inc., under NASA contract NAS 5-03127 for \textit{JWST}. These observations are associated with program \# 5224.

\end{CJK*}

\begin{appendix}

\section{Tests for Systematics}

\begin{figure*}
    \centering
\includegraphics[width=\linewidth]{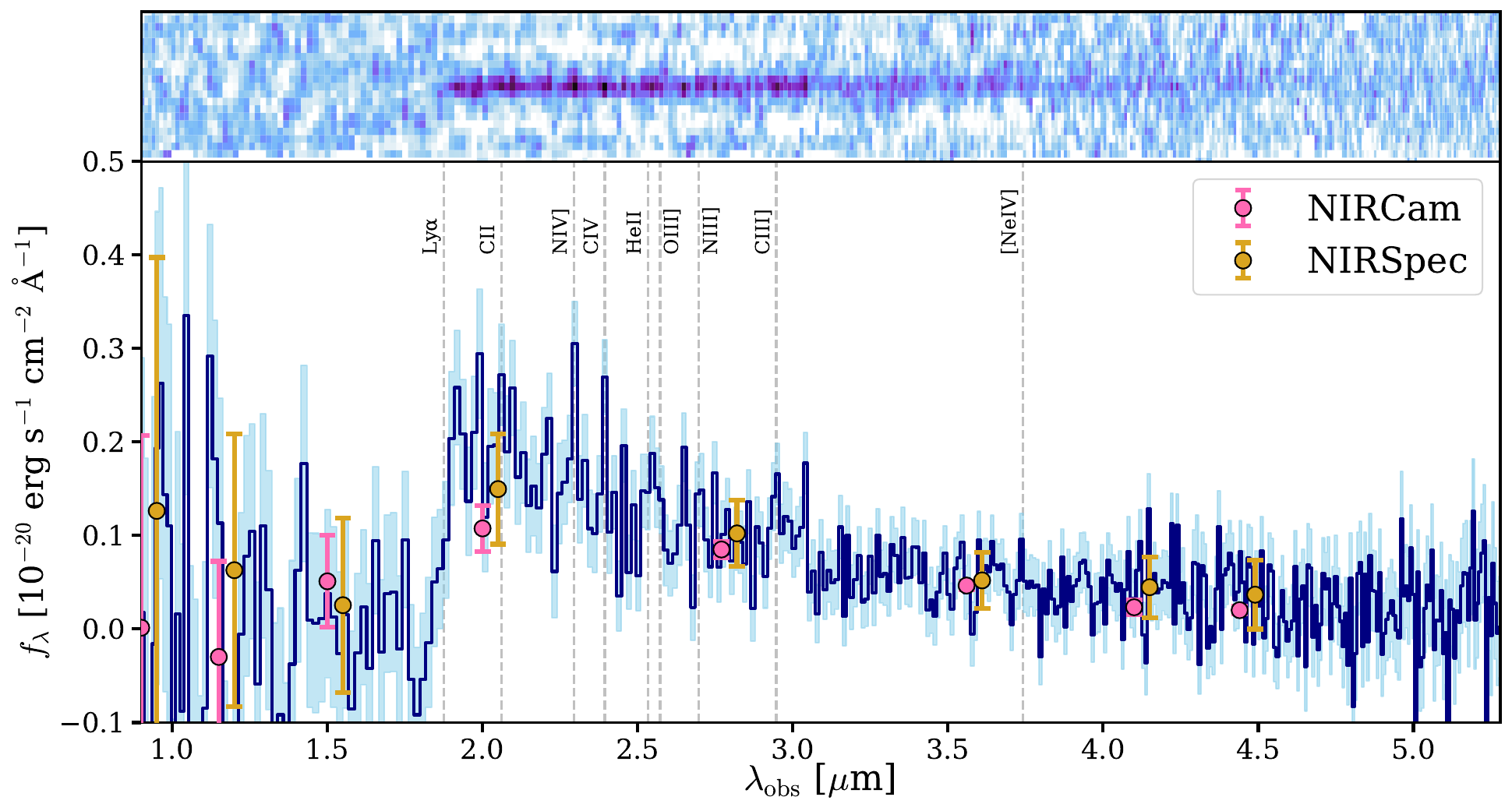}
    \caption{\textbf{Consistency between NIRCam and NIRSpec}. NIRCam fluxes for the source (pink) are compared against synthesized photometry from the spectrum (gold). Points are plotted slightly offset in wavelength for clarity. These fluxes are in excellent agreement across all filters within errors. Note that NIRCam fluxes are not used in any form in the processing of the spectroscopic data with \texttt{msaexp}.}
    \label{fig:nircamvnirspec}
\end{figure*}

We perform two tests to validate different aspects of the NIRSpec data reduction. First, we compare the NIRSpec fluxes to the NIRCam photometry of the source. We synthesize photometry by convolving the observed spectrum with filter curves. The results are shown in Fig. \ref{fig:nircamvnirspec}. We emphasize that at no point in the NIRSpec data reduction are the NIRCam fluxes incorporated (to e.g., rescale the data). In all filters, spanning 0.9-5$\mu$m, the fluxes are in excellent agreement \refrep{($<1\sigma$)} within errors. \refrep{The PRISM/NIRCam ratio of fluxes is F200W: $1.4\pm0.6$; F277W: $1.2\pm0.5$; F356W: $1.1\pm0.7$; F410M: $1.9\pm1.6$; F444W: $1.8\pm1.9$. We note that in all but two bands (F277W, F356W), the NIRCam photometry itself is relatively uncertain ($<5\sigma$). In the two bands (F277W, F356W) where the source is well-detected ($\sim10 \sigma$), the offset is modest ($10-20\%$). Fortunately, these are also the bands whose wavelength coverage spans almost all the key features of interest (e.g., emission lines, $M_{\rm{UV}}$) and so we expect there to be $<20\%$ impact on the derived physical parameters.} This overall consistency we find is crucial to e.g., robustly estimate the $\beta_{\rm{UV}}$ slope or fit the damping wing. 

\begin{figure*}
    \centering
\includegraphics[width=\linewidth]{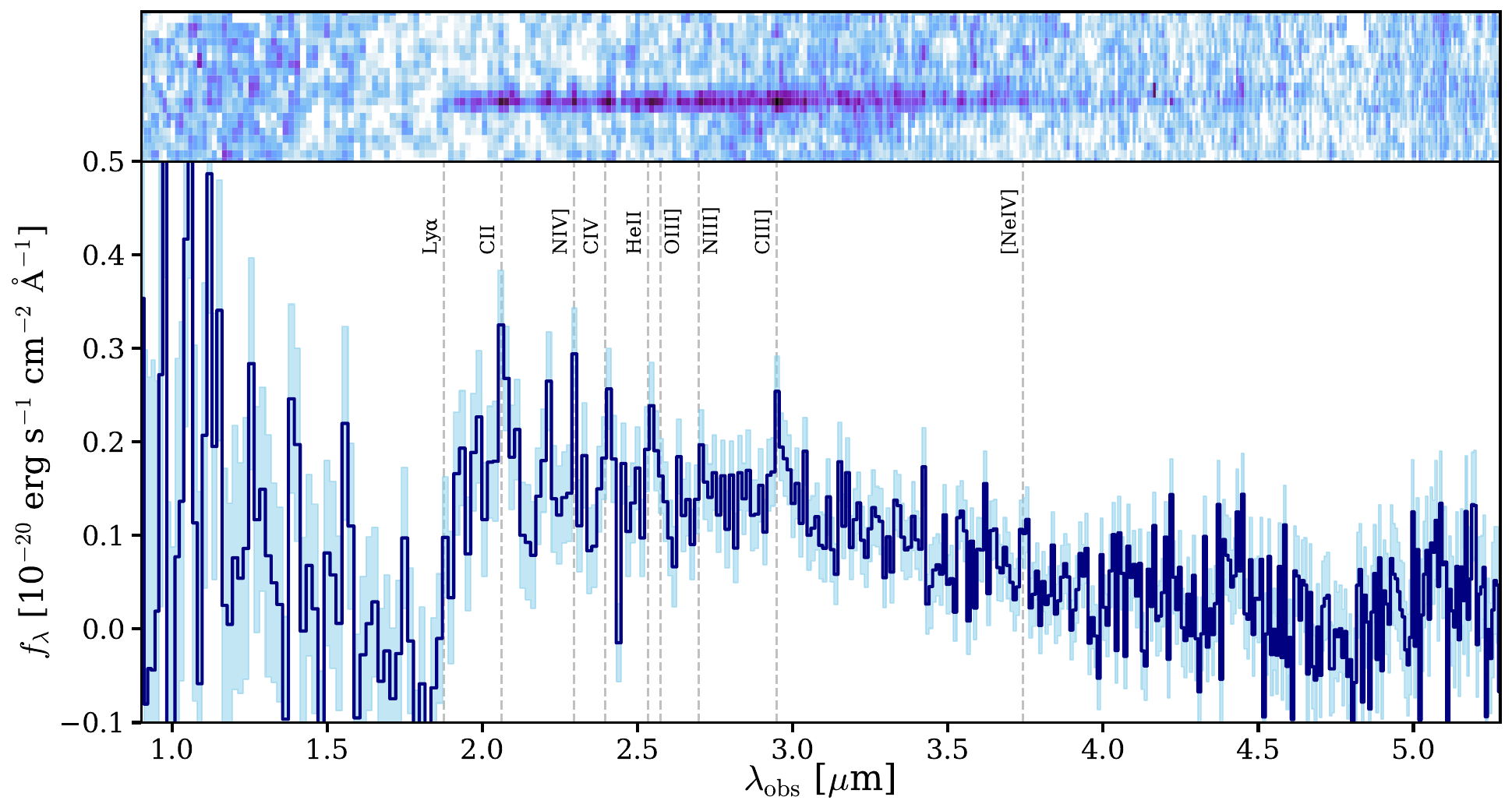}
    \caption{\textbf{Spectrum reduced with a ``global sky" background}. Here we present an alternate way of processing the data -- by constructing a ``global sky" background instead of using the typical ``local" differencing of the 2D spectra that we adopt in our fiducial spectrum. Key features such as the sharp Ly$\alpha$ break and emission lines are recovered at the exact same redshift as in our fiducial reduction. The relative strength and prominence of some lines is somewhat different -- e.g., \nion{C}{iii}] is clearer, and possibly \nion{Ne}{iv} and \nion{C}{ii} are detected at low significance \refrep{($\approx1.4-1.5\sigma$)}.}
    \label{fig:globalsky}
\end{figure*}

In the second test, we consider an alternate form of background subtraction instead of our fiducial ``local" subtraction that makes use of the nod offset positions. In the ``global" background estimation we make use of the empty sky slitlets on the mask that the source was observed in and then fit these data using a sky template with components representing zodiacal dust and a modified Solar spectrum (see Appendix A.2 from \citealt[][]{degraaff24rubies} for details). The resulting spectrum is shown in Fig. \ref{fig:globalsky}. We note that this form of background subtraction comes with its own set of systematics (for e.g., the sky model template may not perfectly track the observed sky). Nonetheless, it provides a handle on the robustness of the features that are the focus of our work. We see that the prominent features from the fiducial spectrum are present here as well, with the various reported lines appearing at $z=14.44$ and the Ly$\alpha$ break consistent with the same redshift.

\section{Ruled Out Low-Redshift Solutions}

\begin{figure}
    \centering
\includegraphics[width=\linewidth]{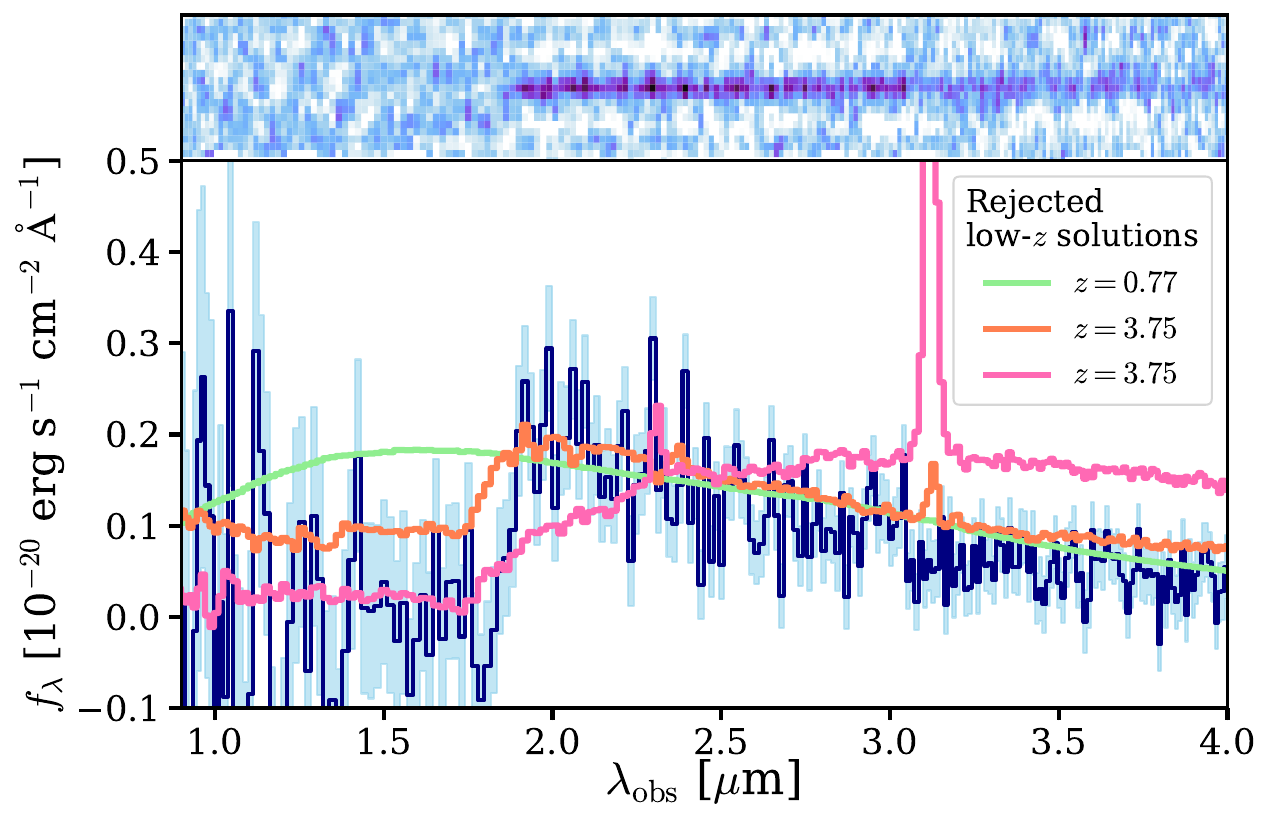}
    \caption{\textbf{Low-$z$ solutions marginally allowed by the NIRCam photometry are completely ruled out with the spectrum.} The two alternate classes of solutions permitted by the photometry (see $p(z)$ in bottom panel of Fig. \ref{fig:specz}) are shown for comparison here -- a quiescent galaxy at $z=3.75$ whose Balmer break occurs around $\approx2\mu$m (orange), and a $z=0.77$ source whose continuum peaks around $\approx2\mu$m and fades below $<1\mu$m (green). 
    Both these solutions, as well as strong emission line contamination \citep{ArrabalHaro23}, are firmly ruled out by the sharpness of the observed break and dearth of flux at shorter wavelengths. \refrep{Additionally, using the spectrum of ``The Cliff" (pink; \citealt[][]{degraaff25}) we demonstrate that ``Black Hole Stars" (BH*s; e.g., \citealt[][]{Naidu25BHstar}) that display deep Balmer breaks cannot explain MoM-z14 as their red optical continuum and broad Balmer lines would be apparent in the data.}}
    \label{fig:lowz}
\end{figure}

One of the key goals of the MoM program was to confirm bright high-redshift galaxy candidates through NIRSpec/prism spectra. In Fig. \ref{fig:lowz}, we show that all the possible low-$z$ solutions that were marginally allowed by the NIRCam photometry are completely ruled out. In particular, the $p(z)$ in bottom panel of Fig. \ref{fig:specz} shows two small peaks: one at $z\sim3.75$ where a Balmer break of a quiescent galaxy at  around $\approx2\mu$m could mimick the Ly$\alpha$ break. Similarly, for a faint $z\sim0.8$ source the continuum can peak at around $\approx2\mu$m before fading below $<1\mu$m.  

The $z=4.9$ galaxy from \citet[][]{degraaff24} is used to illustrate the quiescent case by shifting and resampling the spectrum whereas the best-fit \texttt{Prospector} model that was fit to the NIRCam photometry is shown for the $z=0.77$ scenario. Both these solutions are firmly ruled out by the sharpness of the observed break in the prism spectrum and dearth of flux at shorter wavelengths clearly seen in the 2D spectrum. Furthermore, no lines corresponding to $z=3.75$ appear in the spectrum -- quiescent galaxies at this epoch often show H$\alpha$ in emission due to AGN activity \citep[e.g.,][]{Carnall24}.

\refrep{So-called ``Black Hole Stars" (BH*s; e.g., \citealt[][]{Naidu25BHstar,degraaff25}) that have been proposed to power Little Red Dots \citep[e.g.,][]{Matthee24} display extremely strong, sharp Balmer breaks far exceeding the maximum break strength of stellar populations thereby possibly mimicking Lyman breaks. However, these sources show a red continuum and broad Balmer lines that are ruled out by the data.}

The only remaining possible type contaminant from the photometry $p(z)$ would have been a dusty, strong emission line source whose emission lines could mimick a flat continuum, as seen in the so-called `Schroedinger' galaxy \citep[e.g.,][]{ArrabalHaro23,Naidu22Schro}. However, this is obviously not the case for MoM-z14, and all possible low-$z$ solutions are ruled out.


\begin{figure*}
    \centering
\includegraphics[width=\linewidth]{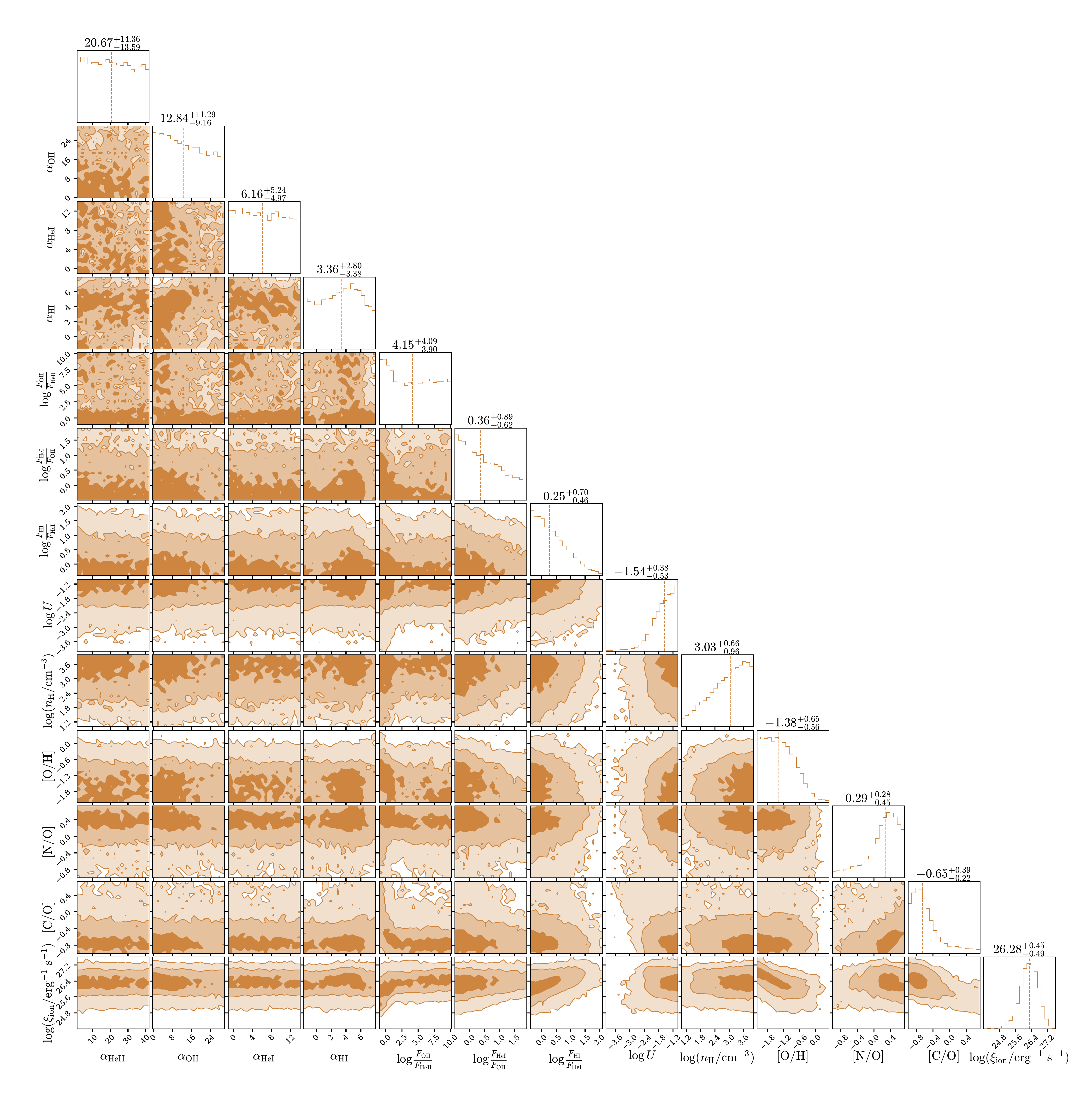}
    \caption{\textbf{Full set of fit \texttt{Cue} parameters}. Same as Fig. \ref{fig:cue}, but here instead of summarizing the ionizing spectrum in the form of $\xi_{\rm{ion}}$, all the power-law slopes ($\alpha$) and normalizations (e.g., $\log \frac{F_\mathrm{OII}}{F_\mathrm{HeII}}$) are shown. The ionizing spectrum parameters are only weakly constrained, but their integral ($\xi_{\rm{ion}}$) implies a strongly ionizing source.}
    \label{fig:cue_appendix}
\end{figure*}

\begin{figure*}
    \centering
\includegraphics[width=0.7\linewidth]{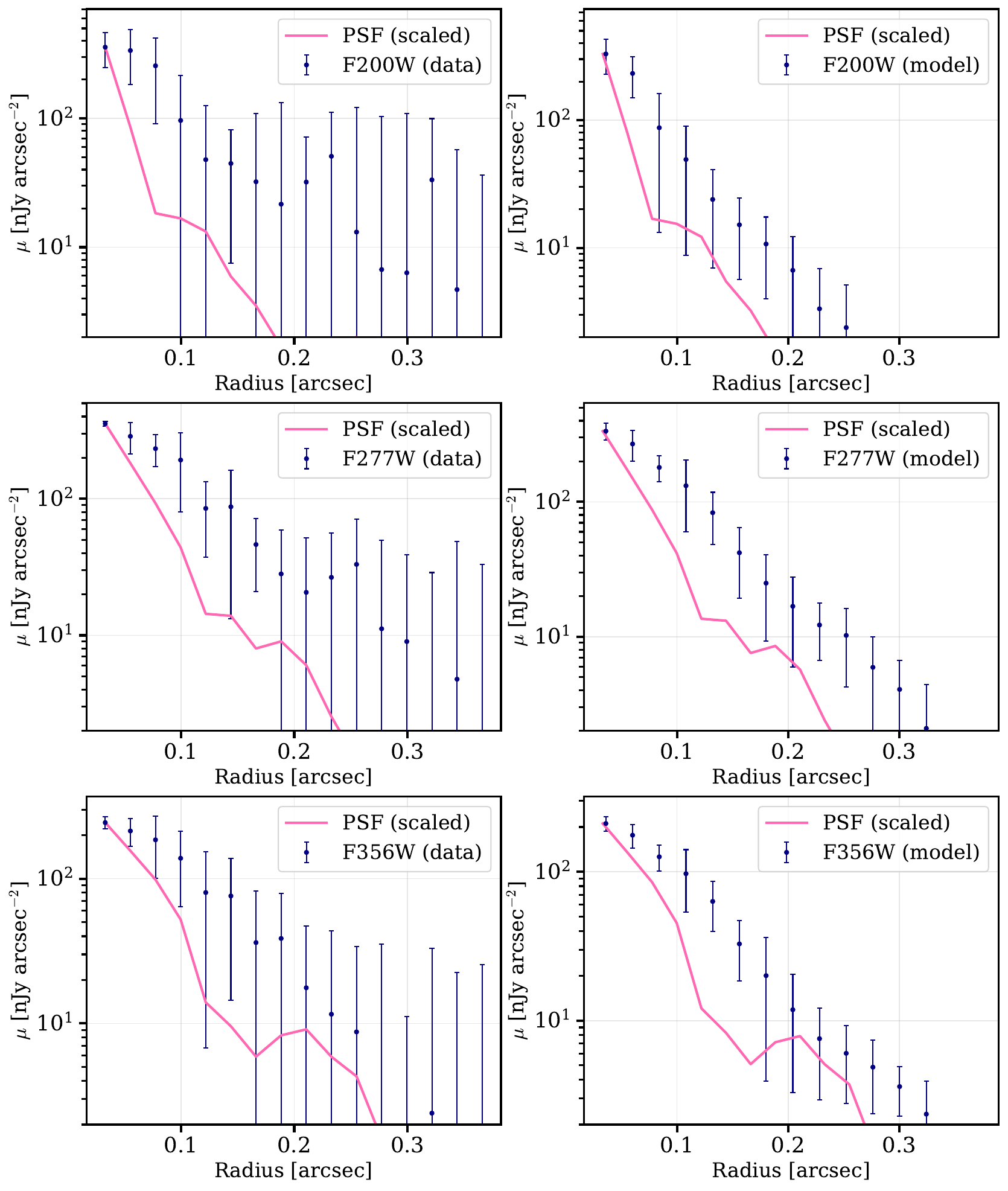}
    \caption{\textbf{Radial surface brightness profiles (navy) compared against the PSF (pink)}. Panels on the left display the data while panels on the right are based on the \texttt{forecepho} model shown in Fig. \ref{fig:morphology}. These profiles that decay gradually relative to the PSF provide independent support that the object is elongated and cannot be described as a point source.}
    \label{fig:radial_profiles}
\end{figure*}

\end{appendix}

\bibliography{MasterBiblio}
\bibliographystyle{apj}

\end{document}